\documentclass[10pt,conference]{IEEEtran}
\usepackage{cite}
\usepackage{amsmath,amssymb,amsfonts}
\usepackage{algorithmic}
\usepackage{graphicx}
\usepackage{textcomp}
\usepackage{xcolor}
\usepackage[hyphens]{url}

\usepackage[hidelinks]{hyperref}
\usepackage{enumitem}
\usepackage{booktabs}
\usepackage{pifont}
\usepackage{color}
\usepackage{tikz}
\usepackage{multirow}
\usepackage{float}
\usepackage{listings}
\usepackage{soul}

\def\BibTeX{{\rm B\kern-.05em{\sc i\kern-.025em b}\kern-.08em
    T\kern-.1667em\lower.7ex\hbox{E}\kern-.125emX}}

\makeatletter
\newcommand{\gnnacc}{FlowGNN}
\newcommand{\newnotecommand}[3]{\newcommand{#1}[1]{\textcolor{#2}{(#3: ##1)}\GenericError{}{Unresolved note by #3 on line \number\inputlineno\ifx\CurrentFile\empty\else\@gobble{} of \CurrentFile\fi}{If this note is resolved, please comment it out or remove it.}{#3: ##1}}}

\newnotecommand{\stf}{cyan}{Stefan}
\newnotecommand{\lks}{teal}{Lakshmi}
\makeatother

\newcommand*\circled[1]{\raisebox{.4pt}
                    {\tikz[baseline=(char.base)]{
            \node[shape=circle,draw,inner sep=1pt, style={fill=black, text=white}, scale=0.75] (char) {\textbf{#1}};}}}
 
\newcommand*\halfcirc[1][0.667ex]{%
  \begin{tikzpicture}
  \draw[fill] (0,0)-- (90:#1) arc (90:270:#1) -- cycle ;
  \draw (0,0) circle (#1);
  \end{tikzpicture}}

\definecolor{codegreen}{rgb}{0,0.6,0}
\definecolor{codegray}{rgb}{0.5,0.5,0.5}
\definecolor{codepurple}{rgb}{0.58,0,0.82}
\definecolor{backcolour}{rgb}{0.95,0.95,0.92}
\lstdefinestyle{codestyle}{
    backgroundcolor=\color{backcolour},   
    commentstyle=\color{codegreen},
    keywordstyle=\color{magenta},
    numberstyle=\tiny\color{codegray},
    stringstyle=\color{codepurple},
    basicstyle=\ttfamily\footnotesize,
    breakatwhitespace=false,         
    breaklines=true,                 
    captionpos=b,                    
    keepspaces=true,                 
    numbers=left,                    
    numbersep=5pt,                  
    showspaces=false,                
    showstringspaces=false,
    showtabs=false,                  
    tabsize=2,
    escapechar=\%,
    columns=fullflexible,
}
\makeatletter
\newcommand\fs@plaintiny{\def\@fs@cfont{\footnotesize}\let\@fs@capt\floatc@plain
\def\@fs@pre{}\def\@fs@post{}%
\def\@fs@mid{\vspace\abovecaptionskip\relax}%
\let\@fs@iftopcapt\iffalse}
\makeatother
\floatstyle{plaintiny}
\newfloat{lstfloat}{htbp}{lop}
\floatname{lstfloat}{Listing}

\title{\huge\gnnacc: A Dataflow Architecture for Real-Time Workload-Agnostic Graph Neural Network Inference} 

\author{Rishov Sarkar, Stefan Abi-Karam, Yuqi He, Lakshmi Sathidevi, Cong Hao\\
School of Electrical and Computer Engineering, Georgia Institute of Technology\\
\{\href{mailto:rishov.sarkar@gatech.edu}{\nolinkurl{rishov.sarkar}}, \href{mailto:stefanabikaram@gatech.edu}{\nolinkurl{stefanabikaram}}, \href{mailto:yhe374@gatech.edu}{\nolinkurl{yhe374}}, \href{mailto:lsathidevi3@gatech.edu}{\nolinkurl{lsathidevi3}}, \href{mailto:callie.hao@gatech.edu}{\nolinkurl{callie.hao}}\}\nolinkurl{@gatech.edu}}

\begin{document}
\maketitle
\thispagestyle{plain}
\pagestyle{plain}

\begin{abstract}

Graph neural networks (GNNs) have recently exploded in popularity thanks to their broad applicability to graph-related problems such as quantum chemistry, drug discovery, and high energy physics.
However, meeting demand for novel GNN models and fast inference simultaneously is challenging due to the gap between developing efficient accelerators and the rapid creation of new GNN models.
Prior art focuses on accelerating specific classes of GNNs, such as Graph Convolutional Networks (GCN), but lacks generality to support a wide range of existing or new GNN models. Furthermore, most works rely on graph pre-processing to exploit data locality, making them unsuitable for real-time applications.
To address these limitations, in this work, we propose a generic dataflow architecture for GNN acceleration, named \textbf{\gnnacc}, which is generalizable to the majority of \textit{message-passing} GNNs.
The contributions are three-fold.
\underline{First}, we propose a novel and scalable \textit{dataflow} architecture, which generally supports a wide range of GNN models with message-passing mechanism. 
The architecture features a configurable dataflow optimized for simultaneous computation of node embedding, edge embedding, and message passing, which is generally applicable to all models. We also propose a rich library of model-specific components. 
\underline{Second}, we deliver ultra-fast real-time GNN inference without \textit{any} graph pre-processing, making it agnostic to dynamically changing graph structures.
\underline{Third}, we verify our architecture on the Xilinx Alveo U50 FPGA board and measure the \textit{on-board end-to-end} performance. We achieve a speed-up of up to 24--254$\times$ against CPU (6226R) and 1.3--477$\times$ against GPU (A6000) (with batch sizes 1 through 1024); we also outperform the SOTA GNN accelerator I-GCN by 1.26$\times$ speedup and 1.55$\times$ energy efficiency over four datasets.
Our implementation code and on-board measurement are publicly available on GitHub.\footnote{\url{https://github.com/sharc-lab/FlowGNN}}

\end{abstract}

\section{Introduction}

Graph Neural Networks (GNNs) have become a powerful tool to apply deep learning to tasks involving graph structures. %
Representative applications of GNNs include analysis of social networks and citation networks, recommendation systems, traffic forecasting, LIDAR point cloud segmentation for autonomous driving, high energy particle physics, molecular representations, and drug discovery~\cite{szklarczyk2019string, wishart2018drugbank, wu2018moleculenet,atz2021geometric}.

\textbf{Real-time Applications.}
GNN inference acceleration is in huge demand, especially for \textbf{real-time} processing.
For instance, point cloud segmentation and detection for autonomous driving using GNNs~\cite{shi2020point} require real-time computation. Another concrete example is in high energy physics (HEP): collision data from a particle collider are collected every 25\textit{ns} and must be processed using GNNs within nanoseconds with raw input graphs to decide whether to save or discard the data; overrunning latency targets will overflow memory buffers and lose precious data~\cite{shlomi2020graph, iiyama2021distance, cerminara2020distance}.
Consequently, we focus on applications with two distinct features %
neglected by previous work.
\underline{First}, \textit{many small graphs are consecutively streamed in at batch size 1 with extremely low latency}~\cite{abcnet,qu2020jet,elabd2022graph}, in contrast to other applications where a single graph is processed.
\underline{Second}, real-time computation must be delivered in a \textit{workload-agnostic} manner, without time for graph pre-processing or graph-specific optimizations.

\textbf{Platform.} FPGAs are desirable for energy-efficient, low-latency inference and have already been used in HEP for GNN acceleration~\cite{elabd2022graph} (however, they only implement one GNN on one hard-coded graph).
Given the rapidly evolving GNN architectures, FPGA provides reconfigurability to quickly adapt to newly proposed GNNs and thus is highly practical.

The \textbf{challenges} and the \textbf{limitations} of existing accelerators, however,
are significant.
\underline{First}, GNN computation is both 
communication- and computation-intensive, as also noted by previous literature~\cite{zhang2020hardware, Li2021meloppr, auten2020hardware}. %
\underline{Second}, novel GNN models are rapidly emerging while accelerator innovation lags behind. Most state-of-the-art (SOTA) GNN accelerators are tailored for Graph Convolutional Networks (GCNs)~\cite{geng2020awb, zhang2020hardware, zhang2021boostgcn, igcn}, which can be conveniently expressed as sparse and general matrix multiplications (SpMM and GEMM). However, complicated operations such as edge embedding, attention, mixed neighborhood aggregation, etc.\ make the majority of GNNs \textit{unsuitable} for SpMM/GEMM. Therefore, to rapidly adapt to evolving GNN models, \textit{generic, extensible, and flexible acceleration frameworks} are needed.
\underline{Third}, some accelerators focus on one large input graph and propose optimizations relying on graph properties. Examples include graph preprocessing to enhance data locality~\cite{zhang2020hardware, jia2020redundancy, wang2020gnnadvisor, chen2021rubik, yan2020hygcn}, and graph partitioning relying on a property of a fixed input graph (e.g., by analyzing adjacency matrix sparsity~\cite{zhang2021boostgcn}).
Such preprocessing or graph-specific techniques \textit{are not feasible} for real-time applications with millions of input graphs with varied structures.

Motivated by the emerging requirements and existing limitations, we propose a generic and flexible architecture for GNN acceleration on FPGA, named \textbf{\gnnacc}, which supports a wide range of prevailing GNNs with message-passing mechanism and is easily extensible for new models.
We summarize our contributions as follows:
\begin{table*}
    \centering
    \footnotesize
    \vspace{-12pt}
    \renewcommand{\arraystretch}{0.9}
    \caption{GNN coverage and features of \textbf{\gnnacc} in comparison with prior works.}
    \vspace{-8pt}
    \setlength{\tabcolsep}{0.5pt}
    \newcommand{\tabcenter}{\centering\let\\\tabularnewline}
    \newcommand{\yes}{\tabcenter\ding{52}}
    \newcommand{\no}{\tabcenter\ding{55}}
    \newcommand{\limited}[1]{\tabcenter\hphantom{\textsuperscript{#1}}\halfcirc\textsuperscript{#1}}
    \begin{tabular}{p{0.125\textwidth}|p{0.04\textwidth}|p{0.04\textwidth}|p{0.04\textwidth}|p{0.04\textwidth}|p{0.04\textwidth}|p{0.04\textwidth}|p{0.1\textwidth}|p{0.1\textwidth}|p{0.1\textwidth}|p{0.1\textwidth}|p{0.1\textwidth}|p{0.1\textwidth}}
        \toprule
        & \multicolumn{6}{c|}{\textbf{Models}} & \multirow{2}{0.1\textwidth}{\centering \textbf{Edge embeddings}} & \multirow{2}{0.1\textwidth}{\centering \textbf{Anisotropic aggregations}} & \multirow{2}{0.1\textwidth}{\centering \textbf{\\Attention}} & \multirow{2}{0.1\textwidth}{\centering \textbf{Flexible dataflow}} & \multirow{2}{0.1\textwidth}{\centering \textbf{%
        Multi-level parallelism*%
        }} & \multirow{2}{0.1\textwidth}{\centering \textbf{No pre-processing}} \\
        \textbf{Accelerator} & \tabcenter GCN & \tabcenter GIN & \tabcenter GAT & \tabcenter PNA & \tabcenter DGN & \tabcenter VN &&&&&& \\
        \midrule
        \textbf{AWB-GCN} \cite{awbgcn} & \yes & \no & \no & \no & \no & \no & \no & \no & \no & \no & \yes & \yes \\
        \textbf{HyGCN} \cite{hygcn} & \yes & \limited1 & \no & \no & \no & \no & \no & \no & \no & \yes & \no & \no \\
        \textbf{I-GCN} \cite{igcn} & \yes & \limited1 & \no & \no & \no & \no & \no & \no & \no & \no & \yes & \yes \\
        \textbf{Auten \textit{et al.}}\ \cite{auten2020hardware} & \yes & \no & \limited2 & \no & \no & \no & \no & \limited3 & \limited2 & \no & \yes & \yes \\
        \textbf{GCoD} \cite{gcod} & \yes & \limited1 & \yes & \no & \no & \no & \no & \limited3 & \yes & \no & \yes & \no \\
        \textbf{ReGNN} \cite{regnn} & \yes & \limited1 & \no & \no & \no & \no & \no & \yes & \no & \no & \yes & \no \\
        \textbf{\textsc{ReFlip}} \cite{huang2022accelerating} & \yes & \limited1 & \yes & \no & \no & \no & \no & \limited3 & \yes & \no & \yes & \yes \\
        \textbf{FlowGNN} & \yes & \yes & \yes & \yes & \yes & \yes & \yes & \yes & \yes & \yes & \yes & \yes \\
        \midrule
        \multicolumn{13}{c}{\ding{52} Full support\qquad\halfcirc\textsuperscript{1} Without edge embeddings\qquad\halfcirc\textsuperscript{2} No attention normalization\qquad\halfcirc\textsuperscript{3} Supports attention only\qquad\ding{55} No support} \\
        \multicolumn{13}{c}{*Supports parallelism both within nodes and between nodes during aggregation (inter- and intra-node parallelism).} \\
        \bottomrule
    \end{tabular}
    \vspace{-12pt}
    \label{tab:existing-list}
\end{table*}
\begin{table}
    \centering
    \footnotesize
    
    \caption{Supported representative GNNs by our framework \textbf{\gnnacc} with flexible extensions.}
    \vspace{-8pt}
    
    \setlength{\tabcolsep}{3pt}
    \begin{tabular}{p{0.095\textwidth}|p{0.335\textwidth}}
    \toprule
    \textbf{Model} & \textbf{Representativeness} \\
    \midrule
    \textbf{GCN} \cite{kipf2016semi} & GNN family that can be represented as sparse matrix-matrix multiplications (SpMM) \\
    
    \hline    
    
    \textbf{GIN} \cite{xu2018powerful,gine} & GNN family with \textit{edge embeddings} and transformations where SpMM \textit{does not} apply  \\
    
    \hline
    
    \textbf{GAT} \cite{velivckovic2017graph} & Anisotropic GNN family with sophisticated message functions\\
    
    \hline
    
    \textbf{PNA} \cite{corso2020principal} & A popular GNN family arbitrarily using multiple aggregation methods \\
    
    \hline
    
    \textbf{DGN} \cite{beani2021directional} & A state-of-the-art GNN with a directional flow at each node and  guided aggregation\\
    
    \hline
    
    \textbf{VN} \cite{gilmer2017neural} & A widely used GNN technique with a virtual node connecting to all other nodes\\
    \midrule
    
    \multicolumn{2}{p{0.45\textwidth}}{\textbf{GCN:} graph convolutional network; \textbf{GIN:} graph isomorphism network; \textbf{GAT:} graph attention network; \textbf{PNA:} principal neighbourhood aggregation; \textbf{DGN:} directional graph network; \textbf{VN:} GNN with virtual node.} \\
    
    \bottomrule
    
    \end{tabular}
    
    \label{tab:GNN-list}
    \vspace{-12pt}
\end{table}
\label{sec:summary-contrib}
\begin{itemize}[leftmargin=*]
    \item {\textbf{Generic.} (1) \textit{Model-generic.} {\gnnacc} is the first generic GNN accelerator able to process the majority of state-of-the-art GNNs by supporting the \textit{message passing} mechanism.
    Table~\ref{tab:GNN-list} summarizes currently supported GNN models, each representative of a large GNN family. In particular, we emphasize {\gnnacc}'s support for \textit{edge embeddings}, which are not considered by existing accelerators but are widely used in most GNN models. (2) \textit{Workload-generic.} {\gnnacc}
    is also dataset and graph structure agnostic; its optimizations do not rely on analysis for specific input graphs, but can effectively process a series of graphs with arbitrary structure.}
    
    \item {\textbf{Real-time.} {\gnnacc} targets \textit{real-time} applications with \textit{zero preprocessing} and partitioning, where the graphs are streamed in consecutively and processed on-the-fly.}
    
    \item {\textbf{Dataflow architecture.} We propose a novel dataflow architecture, which can effectively overlap the two most time-consuming steps in GNN, node transformation and message passing, and significantly reduce processing units' idle time. To boost the performance, FlowGNN also exploits multiple levels of parallelism: node, edge, scatter, and apply, via a novel multi-queue-based on-the-fly multicasting adapter.}

    \item {\textbf{Open-source and modularized.} The implementation of {\gnnacc} is publicly available, with on-FPGA measurement and guaranteed end-to-end functionality by cross-checking with PyTorch code. It has a rich library of GNN modules and an easy-to-use programming model for developing new GNN models inside {\gnnacc} framework.}

    \item {\textbf{Evaluation.} We verify the proposed architecture on Xilinx Alveo U50 FPGA by measuring its
    \textit{on-board} performance. We use seven popular datasets totaling more than 57k graphs being streamed into {\gnnacc}. 
    {\gnnacc} achieves a speed-up of 54--254$\times$ against CPU (6226R) and 1.3--477$\times$ against GPU (A6000) with batch sizes from 1024 to 1 with $4\times$ less power. Comparing with the SOTA accelerator I-GCN, we observe 1.03$\times$ and 1.25$\times$ better performance. The remarkable speedup suggests that \textit{we did not sacrifice performance for generality}, given that the goal of FlowGNN is a general framework for advanced GNNs.}
\end{itemize}

\section{Related Work and Motivations}

\subsection{Related Work}

Recent GNN accelerators are summarized by a survey~\cite{abadal}, including acceleration on CPU/GPUs, ASICs, FPGAs, and heterogeneous platforms.
Auten et al.~\cite{auten2020hardware} and HyGCN~\cite{hygcn} are the earliest ASIC GNN accelerators.
AWB-GCN~\cite{awbgcn} is an FPGA accelerator that aims to combat workload imbalance in graph processing.
EnGN~\cite{engn} uses PEs connected in a ring and performs aggregations using a technique called Ring-Edge Reduce. %
GCNAX \cite{gcnax} addresses resource underutilization and excessive data movement using a flexible dataflow.
Rubik~\cite{rubik} and GraphACT~\cite{graphact} aim to accelerate GCN training using ASIC and FPGA, respectively.
BoostGCN~\cite{zhang2021boostgcn} specifically optimizes GCN via sparsity analysis and graph partitioning.
I-GCN~\cite{igcn} is the most recent GCN accelerator delivering the state-of-the-art performance, which uses an islandization approach to de-duplicate redundant GCN computations by merging nodes with shared neighbors.

\begin{figure}
    \centering
    \vspace{-8pt}
    \includegraphics[width=0.48\textwidth]{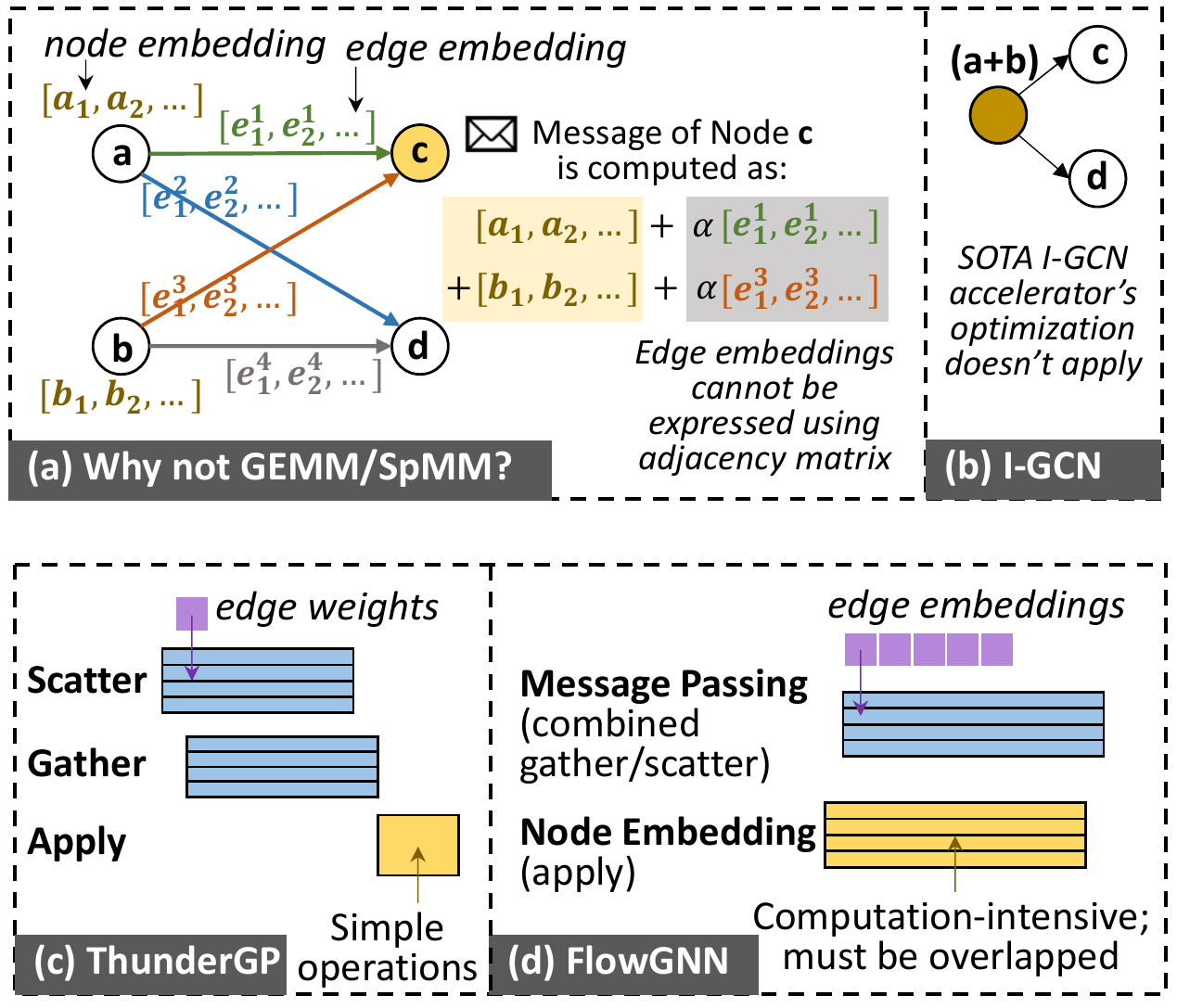}
    \caption{
    (a) GNNs cannot be expressed as a sequence of GEMM and SpMM if there are edge embeddings. (b) With edge embeddings, redundancy removal by node merging in I-GCN~\cite{igcn} is not applicable. (c) and (d) shows the difference between ThunderGP~\cite{chen2021thundergp} and FlowGNN.
    }
    \label{fig:cmp-thunderGP}
    \vspace{-16pt}
\end{figure}

\begin{figure*}
    \centering
    \includegraphics[width=0.95\textwidth]{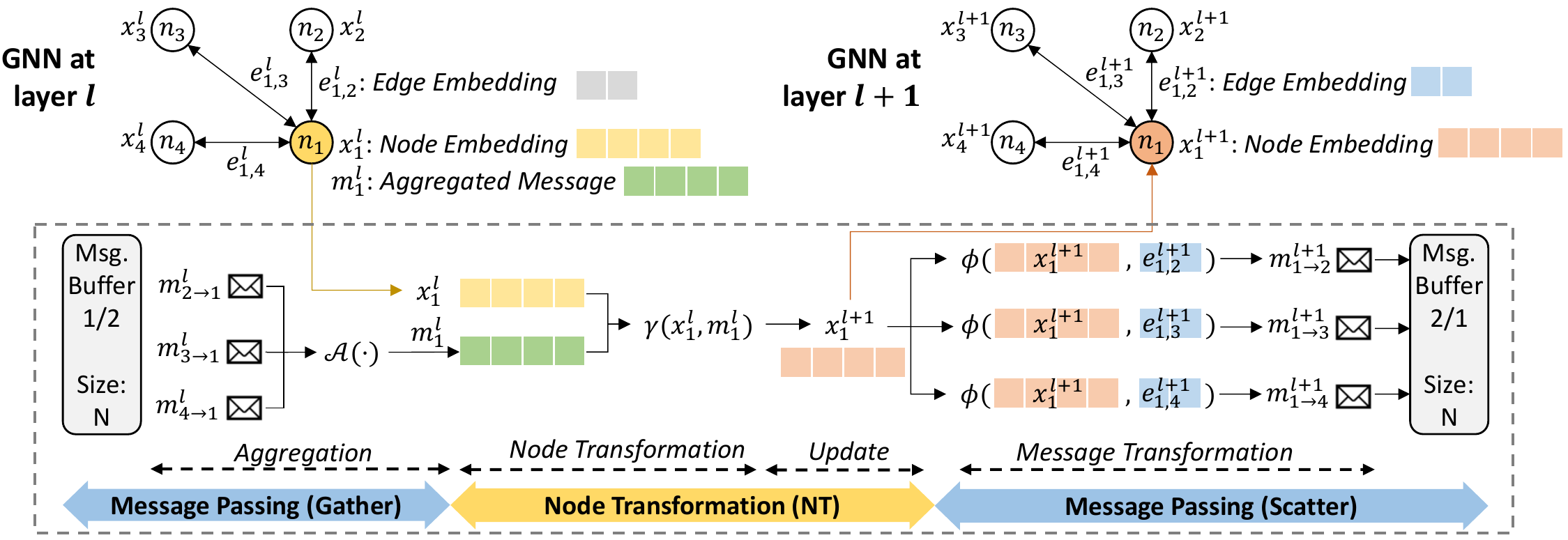}
    \vspace{-12pt}
    \caption{The generic message passing mechanism of prevailing GNN models. It has two major stages: Node Transformation (NT) and Message Passing (MP). NT is analogous to \textit{apply} and MP is analogous to \textit{scatter/gather}. Both NT and MP are usually computation-intensive, especially NT, which typically uses linear or multi-layer perceptron (MLP).}
    \label{fig:gnn-overall}
    \vspace{-12pt}
\end{figure*}

\subsection{Limitations}
\label{sec:limitation}

Despite the great success of existing GNN accelerators and general graph processors, there are still limitations being overlooked. The most significant is that \textbf{advanced GNNs cannot be simplified as matrix multiplications, and edge embeddings cannot be ignored.}
The majority of existing GNN accelerators focus mainly on GCNs and simplify the computation as a series of SpMM and GEMM.
However, this simplification is \textit{not valid} for many reasons. 
\circled{1}
\textbf{Edge embedding.} Edge embeddings are used to represent important edge attributes, such as chemical bonds in a molecule~\cite{yuan2021large, gong2019exploiting} or relationships in social networks, and thus are extensively used in advanced GNNs~\cite{corso2020principal,beani2021directional,gilmer2017neural}. If there are multi-dimensional edge features, the SpMM/GEMM formulation no longer holds. Consider GIN~\cite{gine} with edge embeddings:
\begin{equation}
\textstyle
x_{i}^{l+1}=\text{MLP}((1+\epsilon) \cdot x_{i}^l+\sum_{j \in \mathcal{N}(i)} \text{ReLU}(x_{j}^l+e_{j, i}^l ))
\end{equation}
where $x_{i} \in {X}$ is node embedding and ${e}_{j, i} \in {E}$ is edge embedding. Because the node feature matrix ${X}^{N\times F}$ ($N$~nodes with embedding dimension~$F$) and edge embedding matrix ${E}^{M\times D}$ ($M$~edges with embedding dimension~$D$) differ in size, %
the values have to be added differently. Specifically, the message transformation $\phi(x_j^l,e_{j,i}^l)$ must be computed once for each edge $j,i$, as Fig.~\ref{fig:cmp-thunderGP}(a) depicts. In contrast, without edge embeddings, $\phi(x_j^l)$ can be computed once for each $j$ and reused for all $i$.
\circled{2} \textbf{Invalid existing optimizations.} Edge embeddings make some GNN optimization techniques \textit{unusable}. For example, SOTA GCN accelerator I-GCN~\cite{igcn} proposes redundancy removal by merging nodes with same neighbors to reduce computation, like nodes $a$ and $b$ in Fig.~\ref{fig:cmp-thunderGP}(b). However, with edge embeddings, messages from $a$ to $c$ and $b$ to $c$ now differ from those from $a$ to $d$ and $b$ to $d$, and thus $a$ and $b$ cannot be merged.
\circled{3}
\textbf{Non-trivial aggregation.} GNNs using aggregations besides summation are inexpressible as SpMM. For instance, the Principal Neighborhood Aggregation (PNA) GNN in our framework uses mean, standard deviation, min, and max aggregations altogether in a weighted manner, whose coefficients must be computed on-the-fly and thus disrupts the simple GEMM or SpMM computation pattern.
\circled{4}
\textbf{Anisotropic GNNs.} Anisotropic GNNs require that a neighboring node transformation shall depend on the neighboring node and target node~\cite{tailor2021we}. While for isotropic GNNs, propagation can be implemented using matrix multiplication-style approaches, for anisotropic ones, messages must be explicitly materialized~\cite{tailor2021we}.
One example is Graph Attention Networks (GAT) in our framework. In a GAT layer, a neighboring node is scaled by an attention coefficient $\alpha$, which is dynamically computed based on all neighboring nodes. This dynamic nature prevents GAT from being expressed as matrix multiplications.

Therefore, existing accelerators, which heavily focus on GCN, rely on SpMM/GEMM formulation, and ignore edge embeddings, can greatly limit their generalization to advanced and emerging new GNNs, such as GAT, GIN, PNA, etc.
These advanced GNNs, listed in Table~\ref{tab:GNN-list}, are \textbf{fundamentally different} in their computation because they \textbf{require explicit processing of each edge}; existing SpMM-/\allowbreak{}GEMM-based accelerators \textbf{are unusable} on such models since edge processing will disrupt the GEMM computation patterns; edge embeddings will also invalidate some optimization techniques, e.g., the node redundancy removal in I-GCN~\cite{igcn}.
Meanwhile, many existing accelerators apply graph \textit{pre-processing} to exploit data locality or partitioning on CPU, such as GraphACT~\cite{zeng2020graphact}, HyGCN~\cite{hygcn}, VersaGNN~\cite{shi2021versagnn}, and BoostGCN~\cite{zhang2021boostgcn}. Such pre-processing is usually excluded from performance measurement and can be impractical for real-time applications. %

\subsection{Motivations and Innovations}

Motivated by the above limitations, our goal is to develop a \textit{generic architecture} that can support a wide range of GNNs and can rapidly adapt to emerging GNNs in the future.
The main features of our proposed framework are two-fold: \circled{1} an \textbf{explicit message passing mechanism} that significantly improves the generality of the GNN accelerator for several reasons; \circled{2} \textbf{multi-level parallelism} (inter-node, intra-node, inter-edge across both message passing and node transformation) via multi-queue-based dataflow that can significantly improve performance without losing generality.

\circled{1} \textbf{Explicit message passing for generality.}
First, message passing can express almost all GNN architectures at the theoretical formulation level. As a recent work~\cite{velickovic_2022} states, ``any function of interest we want to compute over graphs can, in all likelihood, be expressed using pairwise message passing.''
Second, explicit message passing can easily integrate edge embeddings and different aggregations without changing the fundamental architecture. We demonstrate the great generality of FlowGNN using six different GNN models (Table~\ref{tab:GNN-list}).

Efficiently supporting message passing without losing generality is non-trivial, however, given the nature of irregular memory access since we can no longer exploit mature SpMM/GEMM optimization techniques. The \textit{real-time} requirement elevates the challenge: we cannot apply any pre-processing to enforce data locality. 

\circled{2} \textbf{Multi-queue multi-level parallelism for performance.}
To address the challenges, our key innovation for high performance is a novel \textbf{multi-queue-based dataflow architecture}.
It can effectively pipeline node transformation and edge embedding, the two most computation-intensive steps in GNN, with multiple levels of parallelism.
In our experiments, our architectural innovation provides all the performance boost with no pre-processing or graph manipulation. Nevertheless, {\gnnacc} is orthogonal to certain optimization techniques such as on-the-fly partitioning and can be applied together.

These innovations make {\gnnacc} a feature-rich framework with excellent GNN model coverage, as shown in Table~\ref{tab:existing-list}. {\gnnacc} fully supports edge embeddings, unlike prior works, as well as anisotropic aggregations and self-attention, which are usually supported incompletely or not at all, with zero pre-processing required. {\gnnacc} also boasts both a flexible dataflow and multi-level parallelism. Existing GEMM-based accelerators exploit both inter- and intra-node parallelism but cannot support general message passing; the only gather/scatter-based accelerator HyGCN does not employ inter-node parallelism in aggregation. {\gnnacc}'s multi-queue-based multicasting enables both intra- and inter-node parallelism within a message passing skeleton.

Since we support explicit message passing, which is similar to the \textit{gather-apply-scatter} (GAS) model in general graph processing, a natural question is how FlowGNN differs from GAS processors.
Fig.~\ref{fig:cmp-thunderGP}(c) and (d) depict the major difference between SOTA graph processor ThunderGP~\cite{chen2021thundergp} and our proposed FlowGNN.
In ThunderGP, the major performance improvement comes from the pipelined \textit{scatter} and \textit{gather}, followed by \textit{apply}. This is acceptable because in graph processing, \textit{apply} usually is a simple operation such as scalar summation. In GNN, however, both message passing (analogous to \textit{gather/scatter}) and node transformation (analogous to \textit{apply}) are computation-intensive, and thus must be effectively pipelined, as we emphasize in FlowGNN.

\section{Generic Architecture}
\label{sec:generic-arch}

In this section, we introduce FlowGNN's generic dataflow architecture, including background of message passing mechanism of GNNs (Sec.~\ref{sec:message-passing-mech}), a baseline simple dataflow design without node-/edge-level parallelism (Sec.~\ref{sec:baseline-flowgnn}), and an improved FlowGNN with multi-level parallelism via data queues (Sec.~\ref{sec:imrp-flowgnn}). Sec.~\ref{sec:model-specific} introduces model-specific components for various GNNs.
Sec.~\ref{sec:programming-model} introduces the programming model.

\subsection{{\gnnacc} Framework Features and Supported GNNs}

\noindent
\textbf{FlowGNN Features.}
Our goal is to provide a real-time GNN acceleration architecture with great generality to support a wide range and emerging GNNs with minimum modifications.
Sec.~\ref{sec:summary-contrib} summarizes the key differentiating features from state-of-the-art accelerators: \textbf{generic}, \textbf{real-time}, and \textbf{open-source}.

\noindent
\textbf{Supported GNNs.}
Table~\ref{tab:GNN-list} summarizes currently supported GNNs, each being representative of a family of GNNs. 
Graph Convolutional Network (GCN)~\cite{kipf2016semi} represents those which can be formulated as SpMM; simplified GCN~\cite{wu2019simplifying} also falls into this category.
Graph Isomorphism Network (GIN)~\cite{xu2018powerful} represents advanced GNNs with higher representation power, \textit{including edge embeddings} and transformations where SpMM \textit{does not} apply; GraphSage~\cite{hamilton2017inductive} falls into this category.
Principal Neighborhood Aggregation (PNA)~\cite{corso2020principal} represents a popular GNN family that uses multiple arbitrary aggregation methods simultaneously.
Graph Attention Network (GAT)~\cite{velivckovic2017graph} represents the anisotropic GNN family with sophisticated message functions, in which edges must be materialized~\cite{tailor2021we}.
Directional Graph Network (DGN)~\cite{beani2021directional} is a state-of-the-art GNN with directional flow at nodes with guided aggregation.
GNN with virtual node (VN)~\cite{gilmer2017neural} is a widely used GNN technique using virtual nodes connected to all other nodes.

\begin{figure*}[t]
    \centering
    \includegraphics[width=0.97\textwidth]{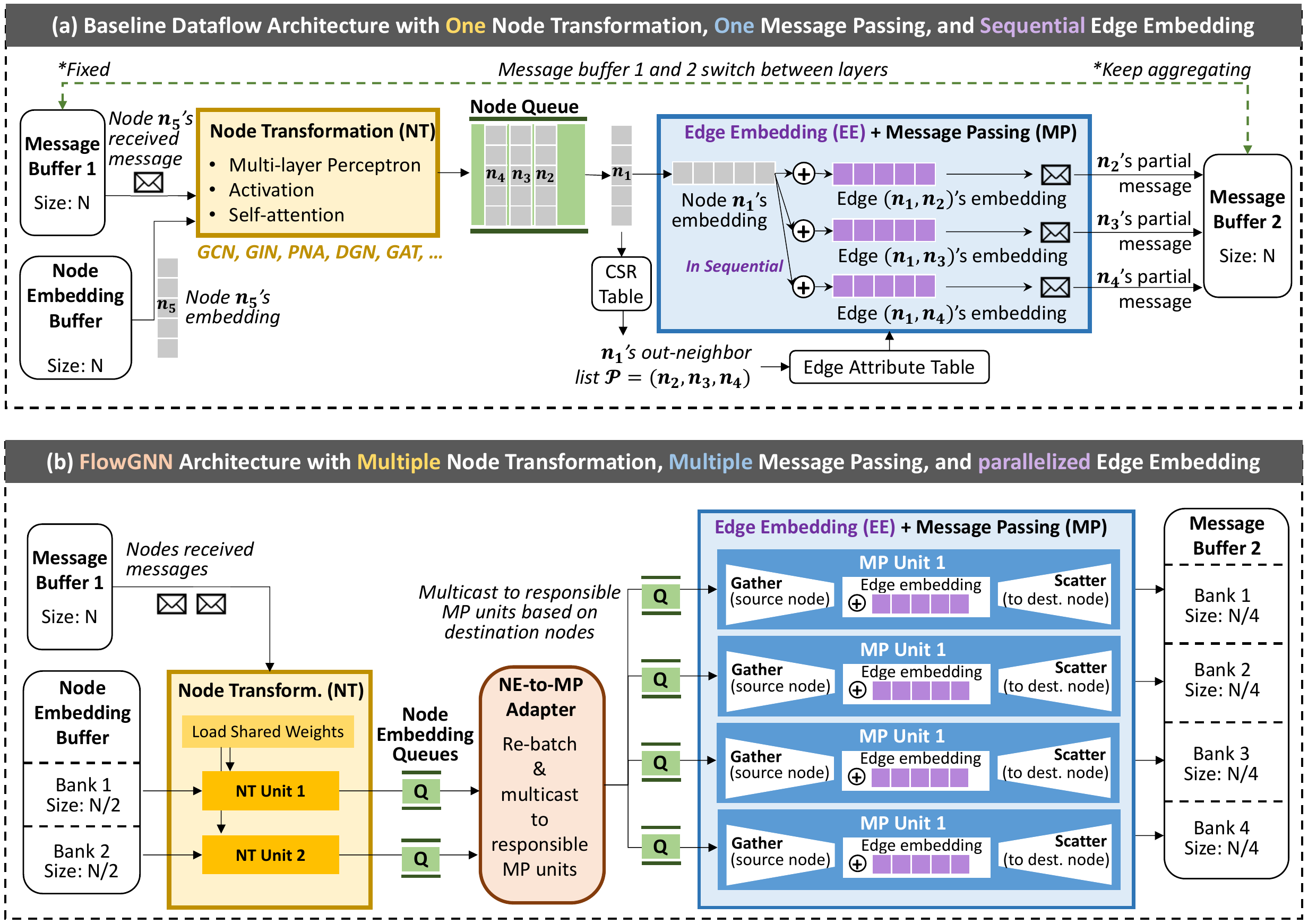}
    \vspace{-12pt}
    \caption{Our proposed baseline dataflow architecture and the improved FlowGNN architecture. (a) The baseline dataflow architecture can effectively pipeline the Node Transformation (NT) and Message Passing (MP), but processes only one node and one edge at a time. (b) The improved FlowGNN architecture can process multiple nodes and multiple edges simultaneously, enabled by an NT-to-MP adapter via on-the-fly multicasting.}
    \label{fig:arch-overall}
    \vspace{-12pt}
\end{figure*}

\subsection{Message Passing Mechanism}
\label{sec:message-passing-mech}

Most prevailing GNN architectures follow the message passing mechanism~\cite{wang2020gcn, gao2019graph, wu2019simplifying, kipf2016semi, xu2018powerful, hamilton2017inductive, corso2020principal, velivckovic2017graph, beani2021directional, gilmer2017neural}.
The general computation of a message passing GNN can be expressed as:
\begin{equation}
\textstyle
x_i^{l+1} = \gamma\Big(x_i^l,\, {\mathcal{A}}_{j\in\mathcal{N}(i)}\big(\phi(x_i^l, x_j^l, e_{i,j}^l)\big)\Big)
\end{equation}
where each node $i$'s embedding $x_i^{l+1}$ at layer $l+1$ is expressed in terms of its embedding $x_i^l$ at layer $l$ along with the embeddings $x_j^l$ of its neighbors $j \in \mathcal{N}(i)$ and their respective edge embeddings $e_{i,j}^l$ through a differentiable message transformation $\phi(\cdot)$, a permutation-invariant aggregation function $\mathcal{A}(\cdot)$, and a differentiable node transformation $\gamma(\cdot)$.

Fig.~\ref{fig:gnn-overall} demonstrates the message passing procedure for a single node $n_1$ at layer $l$, which will be repeated for all nodes and layers.
Highlighted at the bottom of the figure, there are two major steps for each node in each layer: \textbf{message passing (MP)} and \textbf{node transformation (NT)}.
MP is divisible into \textit{gather} and \textit{scatter} phases, where \textit{gather} consists of feature aggregation and \textit{scatter} consists of message transformation and passing. \textit{Edge embeddings} are incorporated into the message during \textit{scatter} phase. NT is usually a linear layer or multi-layer perceptron (MLP), followed by node update.

More specifically, the GNN computation flow has the following stages, as demonstrated in Fig.~\ref{fig:gnn-overall}:

\noindent
\textbf{Message Passing (Gather).}
In the \textit{gather} phase, a.k.a. aggregation, of a certain node $n_1$, the messages from its neighbors obtained in the previous layer are retrieved from a message buffer. The messages are then aggregated in a permutation-invariant manner, denoted by $\mathcal{A}(\cdot)$ (e.g., sum, max, mean, std.\ dev.). In advanced GNNs such as PNA, multiple aggregators are used with learnable weights and scaled based on the degree of the target node. The aggregated message is denoted by $m_{1}^{l}$.

\noindent
\textbf{Node Transformation.}
After aggregation, $m_{1}^{l}$ is processed together with node $n_1$'s current node embedding, denoted by $x_1^l$, via a node transformation function $\gamma(\cdot)$.
This function, with inputs $m_{1}^{l}$ and $x_1^l$, might be
an identity, fully-connected layer, weighted sum, or an MLP.
$\gamma(\cdot)$ produces a new node embedding of $n_1$, denoted by $x_{1}^{l+1}$, and applies the update.

\noindent
\textbf{Message Passing (Scatter).}
After node transformation is the \textit{scatter} phase of message passing. The new node embedding $x_{1}^{l+1}$ will be transformed by a message transformation function $\phi(\cdot)$, usually together with an edge embedding $e_{\text{src}, \text{dest}}^{l+1}$, to generate the node's outgoing messages.
Messages will be dispatched to all neighbors, which will eventually be collected by the \textit{gather} stage of the next layer.%

A complete GNN model may consist of multiple layers, each with message passing and node transformation steps. %
For graph-level tasks, a global pooling layer
is needed, possibly followed by MLP layers for final prediction.

\subsection{Baseline Dataflow Architecture}
\label{sec:baseline-flowgnn}

\begin{figure}[t]
    \centering
    \includegraphics[width=\linewidth]{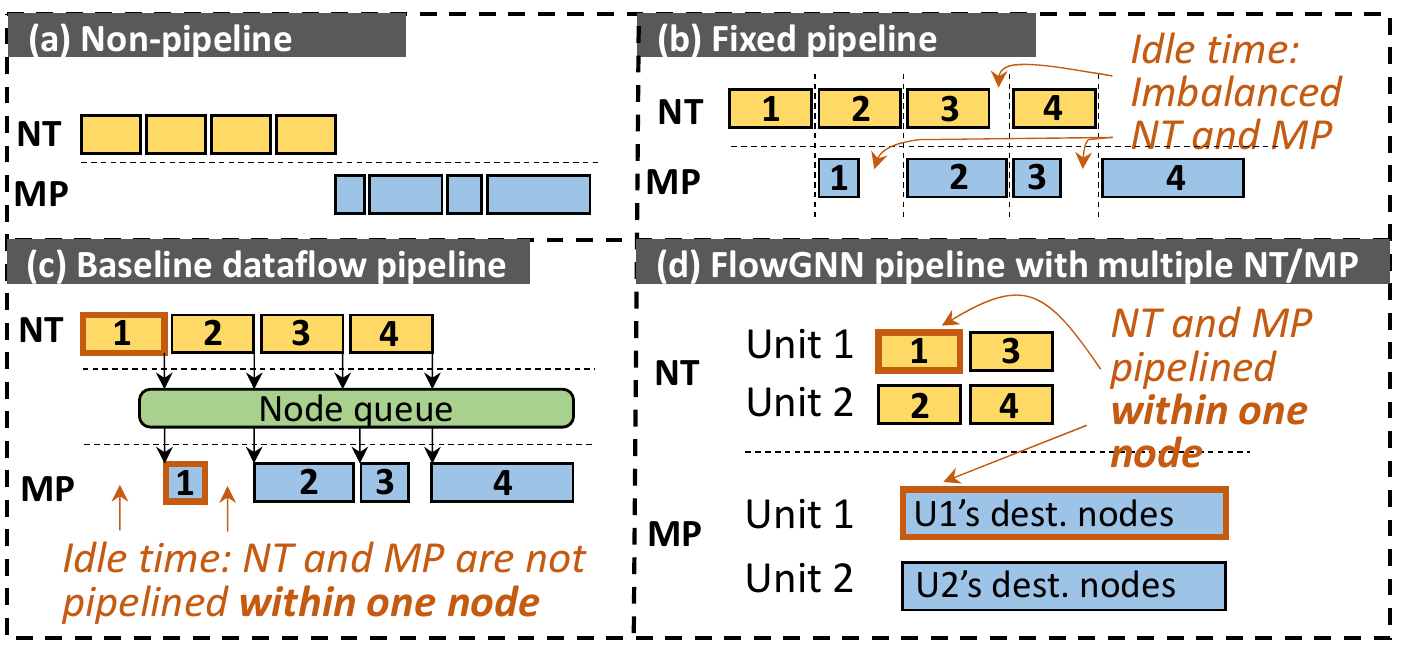}
    \vspace{-20pt}
    \caption{Different strategies of pipelining of node transformation (NT) and message passing (MP). The proposed FlowGNN pipeline in (d) explores node/edge level parallelism and can pipeline NT and MP within one node.}
    \label{fig:pipelining-illustration}
    \vspace{-8pt}
\end{figure}

\begin{figure}[t]
    \centering
    \includegraphics[width=0.48\textwidth]{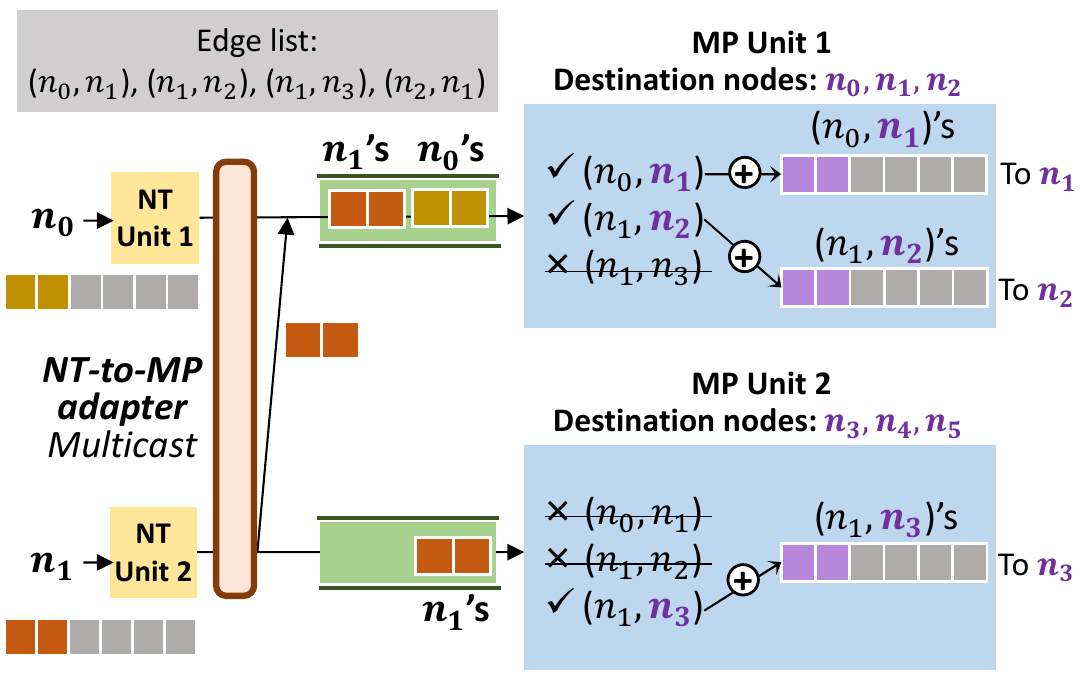}
    \vspace{-8pt}
    \caption{MP units process edges and pass messages based on destination nodes. The NT-to-MP adapter multicasts the node embeddings to the correct MP unit. Note that the MP doesn't wait for the entire node embedding; it starts as soon as the first few elements arrive.}
    \label{fig:filter}
    \vspace{-12pt}
\end{figure}

To explicitly support the message passing mechanism, we first propose the baseline dataflow architecture, shown in Fig.~\ref{fig:arch-overall}(a). 
It has two major processing components: one Node Transformation (NT) unit (yellow block), and one Message Passing (MP) unit (blue block), where edge embeddings are computed during MP.
A node queue between NT and MP, implemented as a FIFO (first-in first-out), holds nodes (their embeddings) that are transformed and ready for message passing.
This queue is the key to enable pipelined NT and MP: as long as the queue is not empty or full, NT and MP can run in parallel.
We keep a high-level abstraction of NT and MP units for now and will introduce more details in Sec.~\ref{sec:nt-mp-unit}.

\noindent
\textbf{Data Buffers.}
The architecture has three data buffers: one node embedding buffer and two message buffers, all with $N$ entries.
The two message buffers act alternately across layers: during layer 1, message buffer 1 is read-only while message buffer 2 is being updated; during layer 2, message buffer 2 becomes read-only and buffer 1 updates, and so on.

\noindent
\textbf{Execution Flow.}
Fig.~\ref{fig:arch-overall}(a) illustrates the execution flow of one GNN layer; for multiple layers, the same resources and dataflow will be reused.
Within one layer,
the NT unit accepts one node, e.g., $n_1$, together with its received message, and applies node transformation and update, e.g., MLP, activation, and self-attention. This is the main component that distinguishes different GNN models.
Then, the MP unit performs the subsequent scatter operation for $n_1$, by sending out messages to all $n_1$'s neighboring nodes.
Consider the graph in Fig.~\ref{fig:gnn-overall} as an example. Once $n_1$'s embedding is updated, the MP unit retrieves all its neighbors, i.e., $n_2$, $n_3$, and $n_4$, computes the messages together with edge embeddings $e_{1,2}$, $e_{1,3}$, and $e_{1,4}$, and dispatches the messages.
The receivers will instantly update their partially aggregated message in the message buffer.
Thus the scatter and gather phases can be merged, since the aggregation function is permutation-invariant, so aggregation order does not matter.
Such a merged fashion has two merits. First, it reduces overall process latency by fusing two stages into one. Second, it reduces memory cost from $O(E)$ to $O(N)$ where $E$ is the number of edges and is typically much larger than $N$.
This approach requires that graph data is stored in its compressed sparse row (CSR) format.

Equivalently, one can first perform gather, i.e., aggregation, via incoming edges, and then node transformation; in this case, no scatter is needed. FlowGNN also supports this dataflow, which is more favorable for GAT. It requires two node embedding and one message buffer, all of size $O(N)$, and graph data stored in compressed sparse column (CSC) format.

Fig.~\ref{fig:pipelining-illustration} illustrates different strategies for NT and MP processing.
\circled{a}~\textbf{Non-pipeline.}
In Fig.~\ref{fig:pipelining-illustration}(a), NT and MP are not pipelined; this apparently incurs a huge waste of idle cycles.
\circled{b}~\textbf{Fixed pipeline.}
In Fig.~\ref{fig:pipelining-illustration}(b), NT and MP are pipelined in a fixed manner: NT for the second node is pipelined with MP for the first node, etc. This reduces some latency but suffers from imbalanced NT and MP processing time. Specifically, if some nodes have larger degrees with longer MP latency than NT while others have shorter MP, there still will be idle cycles.
\circled{c}~\textbf{Baseline dataflow pipeline.}
In Fig.~\ref{fig:pipelining-illustration}(c), NT and MP are pipelined flexibly using a node queue:
as soon as a node finishes its NT, i.e., is ready for message passing, its embeddings are pushed into the queue; meanwhile, the MP unit reads from the queue, incorporates edge embeddings, and then does scatter. This approach can greatly reduce idle cycles.

\noindent
\textbf{Limitations of Baseline Dataflow Architecture.}
Although the baseline dataflow architecture can effectively pipeline NT and MP, there are multiple limitations.
Most obviously, it can process only one node and one edge at a time, restricting an otherwise parallel task.
In addition, as shown in Fig.~\ref{fig:pipelining-illustration}(c), NT and MP are not pipelined within one node, meaning that the node cannot start its MP until its NT completes.

\subsection{Proposed FlowGNN Architecture}
\label{sec:imrp-flowgnn}

Inheriting the advantages of the baseline dataflow architecture, we propose our FlowGNN architecture to address the limitations and to greatly boost performance with multiple levels of parallelism.
FlowGNN allows the following configurable parallelization parameters:
\begin{itemize}[leftmargin=*,noitemsep]
    \item Node parallelism ($P_{\text{node}}$): how many nodes can be processed simultaneously by NT.
    \item Edge parallelism ($P_{\text{edge}}$): how many edges can be processed simultaneously by MP.
    \item Apply parallelism ($P_{\text{apply}}$): how many node embedding dimensions can be processed simultaneously by one NT.
    \item Scatter parallelism ($P_{\text{scatter}}$): how many edge embedding dimensions can be processed simultaneously by one MP.
\end{itemize}

Fig.~\ref{fig:arch-overall}(b) depicts FlowGNN architecture. It has multiple NT units and multiple MP units (2 and 4 in this example, but configurable); between them is an NT-to-MP adapter, connecting NT units and MP units using multiple data queues. To allow parallel access, the node embedding and message buffers are partitioned into multiple banks. We introduce each component in detail in the following.

\subsubsection{Node/Edge Parallelism}
\label{sec:node-edge-parallel}
FlowGNN's key improvement over baseline is node and edge parallelism, which is, however, non-trivial.
While parallelizing NT across multiple nodes is straightforward, it is challenging to parallelize MP across multiple edges, due to the random access of reading edge attributes and reading/writing target node embeddings.
Specifically, to process message passing for a group of edges in parallel, those edges and their target nodes must be in different memory banks or buffers; without graph pre-processing, such accesses are random and thus cannot be parallelized.%

To address this problem, we propose a novel multi-queue dataflow to enable parallelized NT and MP via on-the-fly node distribution.
The key idea is to instantiate multiple NT and MP units, and let each MP unit process its own bank of edges and read/write its own bank of node embeddings. This way, all NT and MP units can operate in parallel with no conflicts.

To achieve this, we design an \textbf{NT-to-MP adapter} and perform \textit{on-the-fly multicasting}.
Given $P_{\text{node}}$ NT units, $P_{\text{node}}$ nodes are processed in parallel as one batch.
During this batch of NT, the NT-to-MP adapter distributes each node's transformed embeddings only to MP units that need to perform any subsequent scatter operations. 
Multicasting is based on \textit{target node IDs}, and edge attributes and node embeddings are also stored in different memory banks based on target node IDs; this ensures that each MP will process only those edges and scatter to only those nodes within in its own bank.
This addresses the first limitation of the baseline dataflow architecture.
Worth noting, for a given node, MP need not wait for its node transformation to complete for all dimensions;
as soon as embedding values are computed, they are streamed into the data queue, which the MP can then fetch. This addresses the second limitation of the baseline architecture.

Fig.~\ref{fig:filter} presents an example of multiple NT and MP units working in parallel via multicasting.
Assume the edge list is $\{(n_0, n_1), (n_1, n_2), (n_1, n_3), (n_2, n_1)\}$.
With two NT units, $n_0$ and $n_1$ are processed in parallel.
As soon as their first $P_{\text{apply}}$ embedding elements are computed (2 in this example), the NT-to-MP adapter will send them to the data queue belonging to the correct MP unit. Assume MP unit 1 is responsible for edges whose destination nodes are in $\{n_0, n_1, n_2\}$; MP unit 2 is responsible for edges with destinations in $\{n_3, n_4, n_5\}$. Since $n_0$'s neighbor is $n_1$, its embeddings should be sent to MP unit 1 only;
since $n_1$'s neighbors are $n_2$ and $n_3$, its embeddings should be sent to both MP units 1 and 2.
Then, each MP unit only processes its own edges in its own memory bank. When the entire embedding is finished, MP unit 1 scatters to nodes $n_1$ and $n_2$, while MP unit 2 scatters to node $n_3$.

Fig.~\ref{fig:pipelining-illustration}(d) shows the \circled{d}~\textbf{FlowGNN pipeline,} a significant improvement over \circled{c}, with multiple NT and MP units executing in parallel, and pipelined NT and MP within one node.

\subsubsection{Apply/Scatter Parallelism}
\label{sec:nt-mp-unit}

\noindent
\textbf{Node Transformation (NT) Unit.}
The NT unit handles any per-node computations required for a GNN. In most GNN algorithms, this is one or more learnable fully-connected layers and activation functions to transform each node's embedding for each GNN layer.

The canonical {\gnnacc} NT unit consists of two sequential processes, overlapped between nodes using ping-pong buffers to hide latency: \textit{accumulate} and \textit{output}. \textit{Accumulate} reads each node's aggregated message and computes a fully-connected layer in an input-stationary fashion: each fetched element of the input vector is used to update the entire output vector. Then, once \textit{accumulate} is finished for an entire node, \textit{output} performs any finalization steps, such as an activation function, and sends the resulting embedding to the multicasting adapter.

The NT unit employs embedding-level parallelism using a configurable parameter $P_{\text{apply}}$, which determines the degree of parallelism across the node embedding dimension. Specifically, it controls how much of a node's message vector is read each cycle to be multiplied to update the output vector. 

\noindent
\textbf{Message Passing (MP) Unit.}
The MP units handle per-edge computations. Two configurations are supported, depending on whether a GNN model is more suited to NT-to-MP (transform, then scatter) or MP-to-NT (gather, then transform) dataflow.

In the NT-to-MP dataflow, each MP unit handles an independent subset of \textit{destination} nodes, as discussed in Sec.~\ref{sec:node-edge-parallel};
analogously, in the MP-to-NT dataflow, each MP unit is assigned a subset of \textit{source} nodes, gathering partial messages along edges from nodes within the assigned subset.

The MP unit also employs a configurable parameter $P_{\text{scatter}}$, determining the degree of parallelism across the edge embedding dimension. Notably, if $P_{\text{apply}}$ and $P_{\text{scatter}}$ are not the same, the NT-to-MP adapter will re-batch the embedding elements for alignment. For example, if $P_{\text{apply}}$ = 1 and $P_{\text{scatter}}$ = 4, the adapter will collect 4 elements from NT and then send to MP.

\section{Model-Specific Components}
\label{sec:model-specific}

\begin{figure}
    \centering
    \includegraphics[width=0.47\textwidth]{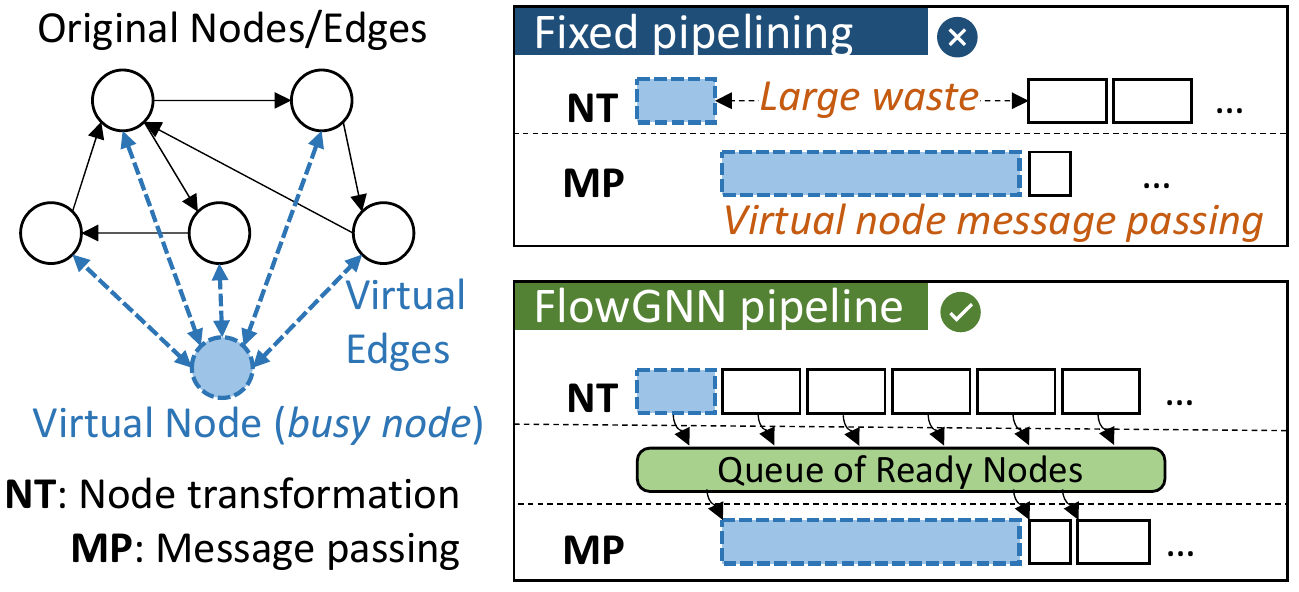}
    \vspace{-12pt}
    \caption{FlowGNN's dataflow architecture is especially beneficial for models with virtual nodes.}
    \label{fig:virtual-node}
    \vspace{-15pt}
\end{figure}

In this section, we introduce model-specific optimizations on top of the general message passing architecture.

\noindent
\textbf{Graph Isomorphism Network.}
GIN is representative of GNNs where message passing involves edge embeddings, and where node transformation is computation-intensive using MLPs.
Its message passing is within the framework using a customized message transformation $\phi(x, m)=x^l + \epsilon^l \cdot m^l$.

\noindent
\textbf{Graph Attention Network.}
GAT is representative of an\-i\-so\-trop\-ic GNNs because of its multi-head self-attention: incoming messages are weighted by attention coefficients based on the node and its neighbor's embeddings. %
GAT is fully compatible with {\gnnacc} through the MP-to-NT dataflow.
GAT uses a custom message transformation %
that weights messages by an attention function %
$A(x_i, x_j)$ such as weighted sum or MLP.

\begin{lstfloat}[t]
\begin{lstlisting}
GNN_compute(graph) {
// GNN compute skeleton; unchanged for all GNNs
  for (layer in model) {
#pragma HLS dataflow // enable queues & pipelining
    node_transformation(nodes_in, nt_out);
    multicasting_adapter(nt_out, mp_in);
    message_passing(mp_in, msg_buf);
    aggregation(msg_buf, nodes_out);
}}
node_transformation(nodes_in, nt_out) {
  for (node in graph) {
    nt_out << new_embedding;
}}
message_passing(mp_in, msg_buf) {
  for (edge in graph) {
    mp_in >> node_data;
}}
aggregation(msg_buf, nodes_out) {
  for (node in graph) {
}}
\end{lstlisting}
\vspace{-11pt}
\caption{\footnotesize FlowGNN pseudocode showing which lines must be modified to create an accelerator for a new GNN.\vspace{-17pt}}
\label{lst:programming-model}
\end{lstfloat}

\noindent
\textbf{Principled Neighbor Aggregation.}
PNA~\cite{corso2020principal} uses multiple neighbor aggregations to increase the distinguishing power:
\begin{equation}
\footnotesize
    \bigoplus = 
        \begin{bmatrix}
            \renewcommand{\arraystretch}{0.8}
            1\\
            \log(D_i+1) / \widetilde{D}\\
            \widetilde{D} / \log(D_i+1)\\
        \end{bmatrix}
    \otimes
        \renewcommand{\arraystretch}{0.8}
        \begin{bmatrix}
            \mu \\
            \sigma \\
            \max \\
            \min
        \end{bmatrix}
\end{equation}
$D_i$ is $x_i$'s degree, and $\widetilde{D}$ is average node degree from training.

PNA falls into the message passing framework with a difference at the aggregation function.
It has four aggregators, min, max, mean, and standard deviation, with scaling values. %

\noindent
\textbf{Directional Graph Networks.}
DGN~\cite{beani2021directional} uses vector fields to define directional flows at nodes for graph convolutions using anisotropic kernels. %
It accepts eigenvectors of the graph Laplacian as parameters %
to compute %
directional aggregation matrices %
during message passing.
DGN uses the mean and directional derivative aggregators as $Y^l = \text{concat}\{ D^{-1}AX^l, |B_{dx}^lX^l| \}$
where ${Y}^{l}$ is aggregated messages, ${X}^{(t)}$ is node embeddings, ${D}$ is the degree matrix, ${A}$ is the adjacency matrix, and ${B}_{dx}^1$ is the directional derivative matrix.
{\gnnacc} is trivially extensible to other DGN aggregations, e.g., directional smoothing~${B}_{av}$. %

\noindent
\textbf{Virtual Node.} \label{sec:virtual-node}
Our dataflow architecture for pipelined node and edge processing especially benefits models with %
virtual nodes~\cite{gilmer2017neural}, artificial nodes connected to all other nodes in the graph. %
They provide shortcuts for
message passing between node pairs and %
are
demonstrated to be effective in many GNN models~\cite{ishiguro2019graph, xue2021node, pham2017graph}.
As shown in Fig.~\ref{fig:virtual-node} left, a virtual node is \textit{busy} with connections to all nodes, resulting in highly unbalanced workloads %
requiring special processing. Some models use multiple virtual nodes~\cite{xue2021node} which escalates the %
complexity.

Fortunately, the proposed dataflow pipelining easily resolves the imbalance introduced by virtual nodes without changing the framework.
As shown in Fig.~\ref{fig:virtual-node} right top, in fixed-pipeline or no-pipeline architectures, virtual node processing will take much longer than other nodes, causing a large waste.
In contrast, Fig.~\ref{fig:virtual-node} right bottom shows how the dataflow architecture can fully overlap virtual node processing with the node embedding computation for other nodes, with zero waste.%

\section{Programming Model}
\label{sec:programming-model}

We now describe the programming model of FlowGNN. Each GNN is compiled to its own kernel and deployed on FPGA. Using FPGA makes it easy to update the accelerator for new state-of-the-art GNN models.

We show the intended workflow for a hypothetical machine learning researcher Alice, who may have some experience with C++ coding but not much with High-Level Synthesis. %
We use pseudocode in Listing~\ref{lst:programming-model} to explain what needs to be changed to adapt to other GNNs in three cases.

First, to try older GNNs such as GraphSage~\cite{hamilton2017inductive}, SGC~\cite{wu2019simplifying}, PAN~\cite{ma2019pan}, UNet~\cite{ronneberger2015u}, or without edge embeddings, Alice can use our GIN, GAT, or PNA kernel without re-compiling and change only the kernel inputs. For example, she may set edge features to all zeros or set a specific aggregator weight to zero.

Second, suppose Alice finds a paper proposing a GNN beyond the current six models, NewGNN. It features an attention mechanism along with min, max, and mean aggregators. As the paper was just released, it has no hardware accelerator yet.

Luckily, FlowGNN already provides modular components with \textbf{extensive coverage for common GNN features.} The message passing skeleton remains unchanged (lines 1--10), and there are already components for attention computation (in our GAT model) and multiple aggregators (in our PNA model). Alice only needs to specify line 6 to be GAT transformation and line 9 to be PNA aggregator to build a NewGNN accelerator.

Third, shortly afterwards, Alice discovers a paper proposing NewerGNN, with novel aggregation and node transformation functions previously unseen in any GNN. Although FlowGNN does not come with built-in components for the new aggregator or node transformation functions, Alice can \textbf{only change a few lines of C++ code} to implement the new features, specifically, the highlighted lines in Listing~\ref{lst:programming-model}. Alice is then able to deploy this accelerator just as quickly and easily as with NewGNN.

\section{Experiments}
\label{sec:results}

\subsection{Model and Implementation Details}
\begin{table}
    \centering
    \caption{Resource usage on Xilinx Alveo U50 FPGA. Clock frequency is 300 MHz.}
    \vspace{-8pt}
    \footnotesize
    \begin{tabular}{c|c|c|c|c}
    \toprule
    \textbf{Model}     &  \textbf{DSP} & \textbf{LUT} & \textbf{FF} & \textbf{BRAM} \\ 
    \midrule
    \textbf{Available} & 5,952 & 872K & 1,743K & 1344  \\\hline
    \textbf{GIN}     & 1,741 & 262,863 & 166,098 & 204 \\
    \textbf{GCN}     & 1,048 & 229,521 & 192,328 & 185 \\
    \textbf{PNA}     & 2,499 & 205,641 & 203,125 & 767 \\
    \textbf{GAT}     & 2,488 & 148,750 & 134,439 & 335 \\
    \textbf{DGN}     & 1,563 & 200,602 & 156,681 & 462 \\
         
    \bottomrule
    \end{tabular}
    \label{tab:resource}
    \vspace{-10pt}
\end{table}
We deploy {\gnnacc} for each of six models using High-Level Synthesis (HLS) by Vitis HLS and Vivado for the Xilinx FPGA Alveo U50 accelerator card, whose available resources are shown in Table~\ref{tab:resource}, targeting a 300 MHz clock frequency.
Graphs are consecutively streamed into the accelerator in raw edge-list format (i.e., COO) with zero CPU intervention.

As listed in Table~\ref{tab:GNN-list}, we implement six GNN models, each representative of a family of GNNs.
Each model has a PyTorch implementation, to which we cross-check our on-board implementation to verify our end-to-end execution is correct.
For GCN, GIN, and GIN-VN, we use 5 layers and node embedding dimension 100, mirroring the PyTorch models~\cite{ogb-models}, along with global average pooling and an output head with one linear layer.
For PNA, we use 4 layers with node embedding dimension 80, global average pooling, and an MLP-ReLU head of sizes (40, 20, 1).
For DGN, we use 4 layers and node embedding dimension 100, global average pooling, and an MLP-ReLU head of sizes (50, 25, 1).
For GAT, we use 5 layers with 4 heads and 16 features each, global average pooling, and an output head with one linear layer.
We use two NT units and four MP units for all models.
Table~\ref{tab:resource} reports each model's resource utilization.

\subsection{Dataset and Baseline}

\begin{table}
    \centering
    \caption{Number of graphs, (average) nodes, and (average) edges in each dataset used for evaluation. We also indicate datasets that include edge features (EF).}
\vspace{-8pt}
    \footnotesize
    \begin{tabular}{c|c|c|c|c}
        \toprule
        \textbf{Dataset} & \textbf{Graphs} & \textbf{Nodes} & \textbf{Edges} & \textbf{EF} \\
        \midrule
        \textbf{MolHIV} & 4113 & 25.3 & 55.6 & \ding{52} \\
        \textbf{MolPCBA} & 43773 & 27.0 & 59.3 & \ding{52} \\
        \textbf{HEP} & 10000 & 49.1 & 785.3 & \ding{52} \\
        \midrule
        \textbf{Cora} & 1 & 2708 & 5429 & \ding{55} \\
        \textbf{CiteSeer} & 1 & 3327 & 4732 & \ding{55} \\
        \textbf{PubMed} & 1 & 19,717 & 44,338 & \ding{55} \\
        \textbf{Reddit} & 1 & 232,965 & 114,615,892 & \ding{55} \\
        \bottomrule
    \end{tabular}
    \label{tab:datasets}
    \vspace{-10pt}
\end{table}

\noindent
\textbf{Datasets.}
Table~\ref{tab:datasets} lists the seven datasets used for evaluation.
We first adopt MolHIV and MolPCBA, two molecular property prediction datasets from the Open Graph Benchmark~\cite{hu2020ogb}, as well as 10k graphs from High Energy Physics (HEP)~\cite{kasieczka2019top}, generated according to the EdgeConv method~\cite{qu2020jet,duarte2020graph,shlomi2020graph} with $k$ = 16.
These datasets model our target use case of real-time inference on many small graphs (10--100 nodes) streamed in; similar applications have been demonstrated in other GNN works \cite{streamgcn,qu2020jet}.
Moreover, these datasets include edge features for comprehensive evaluation of GNNs with edge embeddings.
We also use four prevailing single-graph benchmarks, CiteSeer, Cora, PubMed, and Reddit~\cite{sen2008collective}.

\noindent
\textbf{Baselines.}
As the baseline, we average five iterations of the time measured on CPU (Intel Xeon Gold 6226R) and GPU (NVIDIA RTX A6000), where each model is implemented in PyTorch Geometric~\cite{pyg}. %
On GPU, we use batch sizes from 1 to 1024. Note that batch size 1 is the \textit{only fair comparison} for real-time inference of streaming graphs---batching graphs delays their processing. Other works~\cite{qu2020jet,groqinc2020challenge} also commonly compare only with batch size 1; we provide results at higher batch sizes only to affirm FlowGNN's advantage.

We also compare with the state-of-the-art accelerator, I-GCN~\cite{igcn}, following their experiment settings.
Worth noting, as discussed in Sec.~\ref{sec:limitation}, I-GCN employs a redundancy removal technique that skips explicit materialization of groups of edges, which limits I-GCN's applicability to a narrow class of \textit{isotropic} GNNs with \textit{no edge embeddings}; thus this is \textit{not a fair comparison} since I-GCN will fail in many cases where FlowGNN can manage. Nevertheless, we still make effort to make a comparison.
For other GNN accelerators such as \cite{zhang2020hardware, zhang2021boostgcn}, we find it hard to figure the GNN details such as the number of layers, node embedding dimensions, and whether graph/weight loading and graph-level processing are included. %
Therefore, our main comparison is with CPU and GPU using on-board measured performance with functionality guarantee.

\subsection{End-to-end Evaluation}
\begin{figure*}
    \centering
    \includegraphics[width=\linewidth]{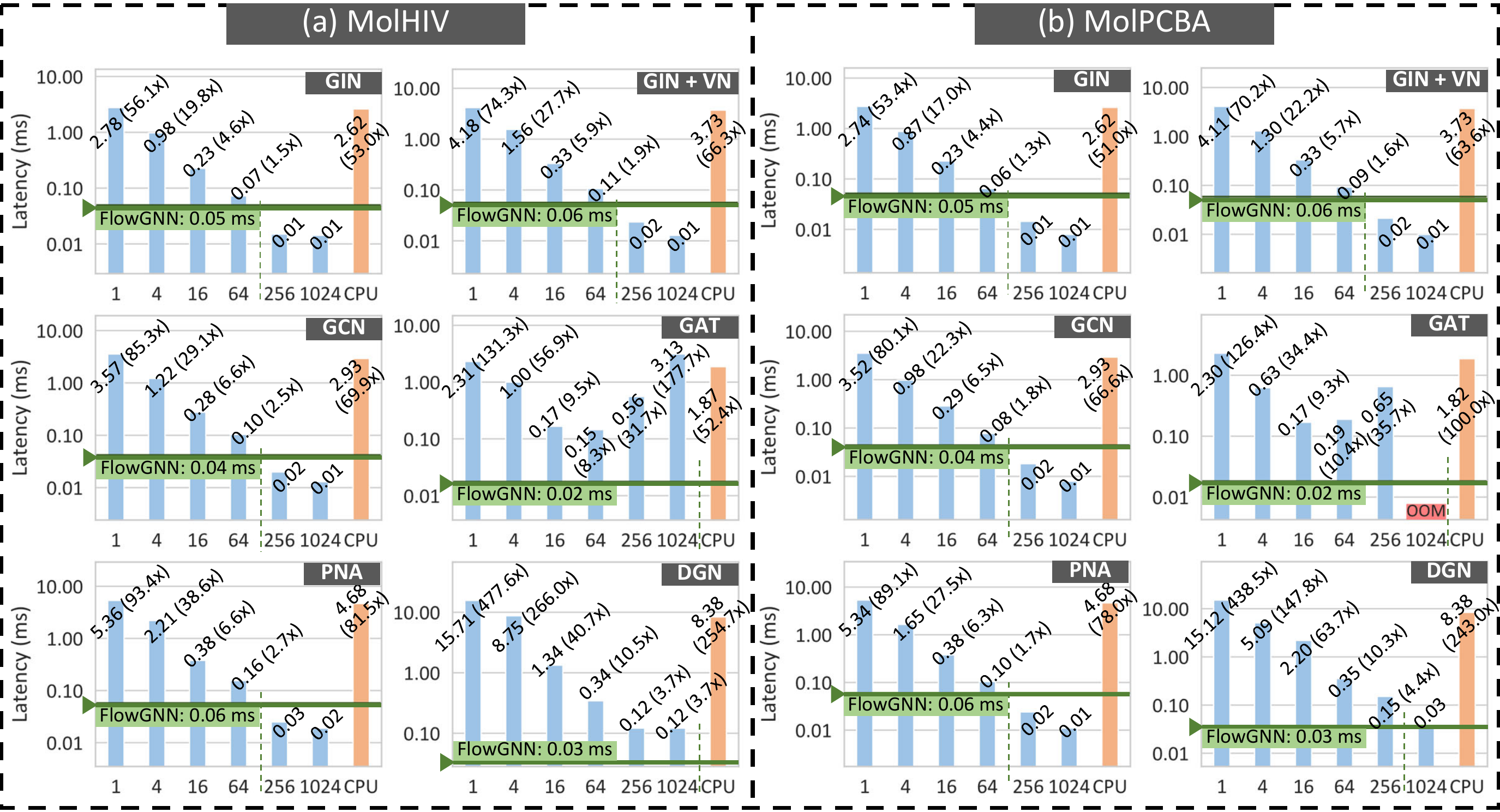}
    \vspace{-20pt}
    \caption{FlowGNN latency measured on-board averaged from 40k test graphs. $x$-axis is GPU/CPU batch size; $y$-axis is average latency (ms) per graph. GPU baseline is evaluated with batch sizes from 1 to 1024. CPU is evaluated at batch size 1. (a) MolHIV results. (b) MolPCBA results. Our FlowGNN outperforms GPU by  3.58--477$\times$ and is consistently faster than GPU with batch size $\leq$ 64; the GAT and DGN models even outperform GPU under batch size 1024.}
    \label{fig:latency_stats}
    \vspace{-10pt}
\end{figure*}

\begin{table}
    \centering
    \caption{On-board latency (in ms) of FlowGNN compared against CPU and GPU at batch size 1, averaged from 10k graphs from high energy physics.}
    \vspace{-8pt}
    \footnotesize
    \renewcommand{\arraystretch}{0.95}
    \begin{tabular}{c|c|c|c}
        \toprule
        \textbf{Model} & \textbf{CPU} & \textbf{GPU} & \textbf{FlowGNN (vs.\ GPU)} \\
        \midrule
        \textbf{GIN} & 4.23 & 2.38 & \textbf{0.1799 (13.3$\times$)} \\
        \textbf{GIN+VN} & 5.02 & 3.51 & \textbf{0.2076 (16.9$\times$)} \\
        \textbf{GCN} & 4.59 & 3.01 & \textbf{0.1639 (18.4$\times$)} \\
        \textbf{GAT} & 2.24 & 1.96 & \textbf{0.0544 (36.0$\times$)} \\
        \textbf{PNA} & 9.66 & 5.37 & \textbf{0.1578 (34.0$\times$)} \\
        \textbf{DGN} & 30.20 & 61.26 & \textbf{0.1382 (443.4$\times$)} \\
        \bottomrule
    \end{tabular}
    \label{tab:hep}
\end{table}

\begin{table}
    \vspace{-8pt}
    \centering
    \caption{Energy efficiency (in graphs/kJ) of FlowGNN compared against CPU and GPU at batch size 1. The MolHIV dataset is used for energy evaluation.%
    }
    \vspace{-8pt}
    \renewcommand{\arraystretch}{0.95}
    \footnotesize
    \begin{tabular}{c|c|c|c}
        \toprule
        \textbf{Model} & \textbf{CPU} & \textbf{GPU} & \textbf{FlowGNN (vs.\ GPU)} \\
        \midrule
        \textbf{GIN} & 4.48{\scriptsize E}3 & 4.50{\scriptsize E}3 & \textbf{7.34{\scriptsize E}5 (163$\times$)} \\
        \textbf{GIN+VN} & 3.16{\scriptsize E}3 & 2.99{\scriptsize E}3 & \textbf{6.46{\scriptsize E}5 (216$\times$)} \\
        \textbf{GCN} & 4.02{\scriptsize E}3 & 3.50{\scriptsize E}3 & \textbf{8.88{\scriptsize E}5 (254$\times$)} \\
        \textbf{GAT} & 6.29{\scriptsize E}3 & 5.41{\scriptsize E}3 & \textbf{2.29{\scriptsize E}6 (424$\times$)} \\
        \textbf{PNA} & 2.52{\scriptsize E}3 & 2.33{\scriptsize E}3 & \textbf{6.11{\scriptsize E}5 (262$\times$)} \\
        \textbf{DGN} & 1.40{\scriptsize E}3 & 7.96{\scriptsize E}2 & \textbf{1.39{\scriptsize E}6 (1748$\times$)} \\
        \bottomrule
    \end{tabular}
    \label{tab:energy}
    \vspace{-10pt}
\end{table}

We fully evaluate {\gnnacc} implementations on six GNN models after place-and-route against CPU and GPU baselines. We measure latency \textbf{on-board end-to-end}, including weight loading and graph loading, and average across the test graphs in each dataset. We report speedups for each batch size.

Results are depicted in Fig.~\ref{fig:latency_stats}(a) for the MolHIV dataset and Fig.~\ref{fig:latency_stats}(b) for the MolPCBA dataset.
In each figure, $x$-axis is GPU/CPU batch size; $y$-axis is the average latency per graph in milliseconds. GPU baseline is evaluated with batch sizes from 1 to 1024, and CPU is evaluated at batch size 1.

It clearly shows that, for six GNN models, on both datasets, {\gnnacc} achieves remarkable speedup over CPU and GPU.
Compared with CPU, FlowGNN is 51.0--254.7$\times$ faster.
Compared with GPU, FlowGNN is 53.4--477.6$\times$ faster at batch size 1. Additionally, FlowGNN consistently outperforms GPU at batch sizes up to 64, from 1.3$\times$ to 266.0$\times$.
The GAT and DGN models even surpass GPU at batch sizes 256 and 1024: GAT is 31.7--177.7$\times$ faster, and DGN is 3.7--4.4$\times$ faster.

Additionally, Table~\ref{tab:hep} lists results on our HEP dataset; all platforms use batch size 1 to simulate real-time high energy physics. FlowGNN is 13.3--443.4$\times$ faster than GPU and 23.5--218.6$\times$ faster than CPU, confirming its real-time advantage.

\begin{figure}[t]
    \centering
     \vspace{-8pt}
    \includegraphics[width=\linewidth]{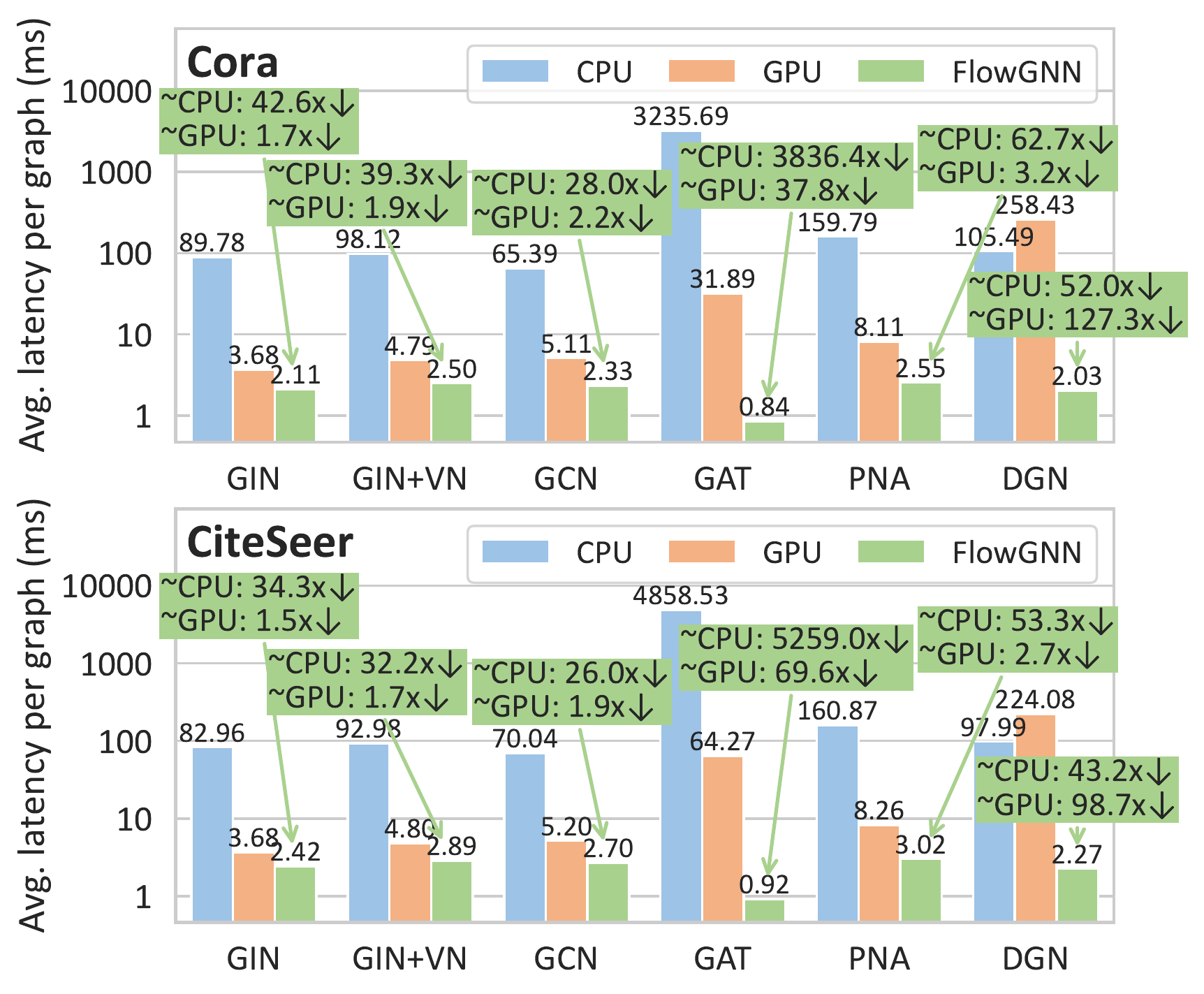}
    \vspace{-24pt}
    \caption{FlowGNN latency on the Cora and CiteSeer datasets, compared against CPU and GPU baselines.}
    \label{fig:cora_citeseer}
    \vspace{-12pt}
\end{figure}

We also evaluated FlowGNN on six GNN models on two popular GNN datasets, Cora and CiteSeer; results are in Fig.~\ref{fig:cora_citeseer}. On both graphs, FlowGNN consistently outperforms CPU and GPU (batch size 1 since these are single graphs). %
GAT and DGN see especially massive speedups of 37.8$\times$ and 127.4$\times$ over GPU on Cora and 69.6$\times$ and 98.7$\times$ on CiteSeer.

Finally, we evaluate FlowGNN's energy efficiency against CPU and GPU on the six models on the MolHIV dataset at batch size 1. 
Results in Table~\ref{tab:energy} show 164--991$\times$ energy efficiency over CPU and 163--1748$\times$ over GPU.

The remarkable speedup and energy efficiency demonstrate the effectiveness of our proposed dataflow architecture, especially with no pre-processing. Although FlowGNN's goal is to be a \textit{generic} framework, the superior results suggest that \textit{we did not sacrifice performance for generality}.

\subsection{Ablation Study}

\begin{figure}
    \centering
    \includegraphics[width=0.44\textwidth]{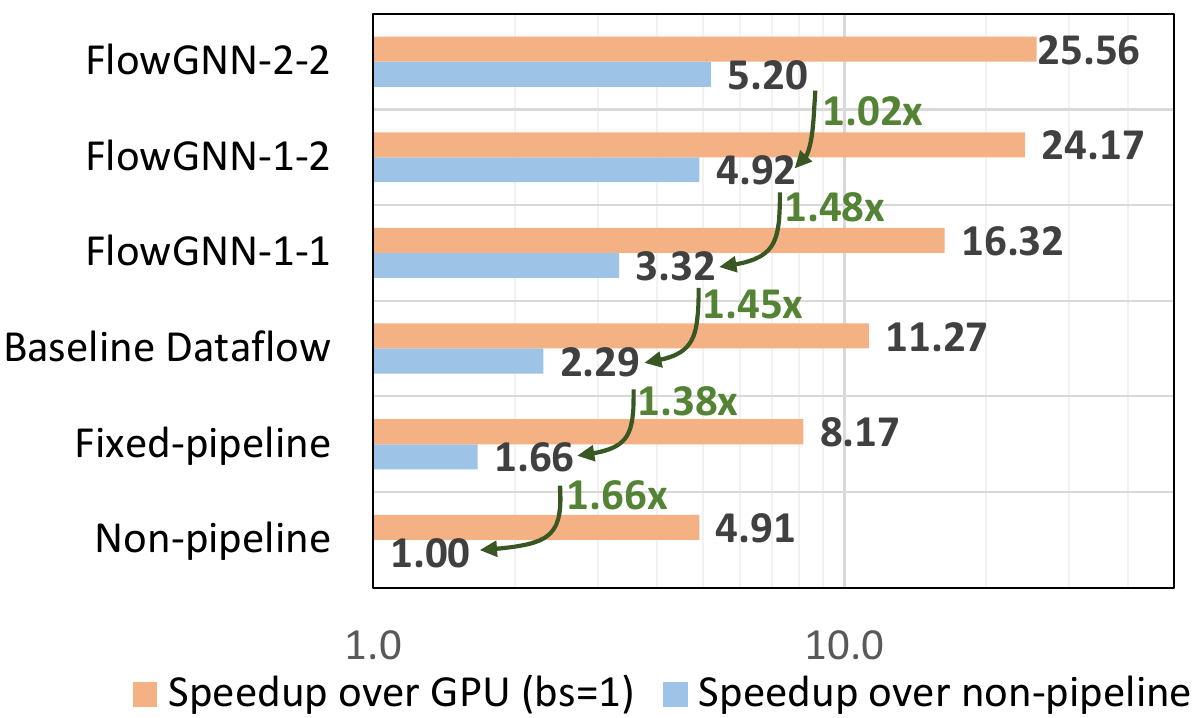}
    \vspace{-8pt}
    \caption{Effectiveness of FlowGNN's dataflow architecture. Speedup plotted in log-scale. FlowGNN results are in the format of FlowGNN-$P_{\text{apply}}$-$P_{\text{scatter}}$.}
    \label{fig:ablation}
    \vspace{-12pt}
\end{figure}

We conduct an ablation study to quantitatively evaluate the baseline and FlowGNN architectures, using the GCN model on MolHIV dataset.
Fig.~\ref{fig:ablation} shows the incremental improvements over the most naive framework, non-pipelined NT and MP (Fig.~\ref{fig:pipelining-illustration}(a), Sec.~\ref{sec:baseline-flowgnn}).
It shows that thanks to our customized NT and MP units with default parallelism (e.g., parallel linear and MLP), the non-pipelined NT and MP scheme is already 4.91$\times$ faster than GPU. Next, it shows that fixed-pipeline is 1.66$\times$ faster than non-pipeline, while the baseline dataflow achieves another 1.38$\times$ speedup by reducing NT and MP idle cycles.
Then, FlowGNN-1-1 ($P_{\text{apply}}$ = 1, $P_{\text{scatter}}$ = 1) is 1.45$\times$ faster than the baseline dataflow, since it pipelines NT and MP for one node: MP can start before the entire NT is finished. We further increase the $P_{\text{scatter}}$ from 1 to 2 and $P_{\text{apply}}$ from 1 to 2, leading to 1.48$\times$ and 1.02$\times$ improvement, respectively.
This ablation study shows that our proposed baseline dataflow architecture, along with pipelined NT/MP within one node and multiple NT/MP, are very effective in reducing the latency.

\begin{figure}[t]
    \centering
    \includegraphics[width=0.92\linewidth]{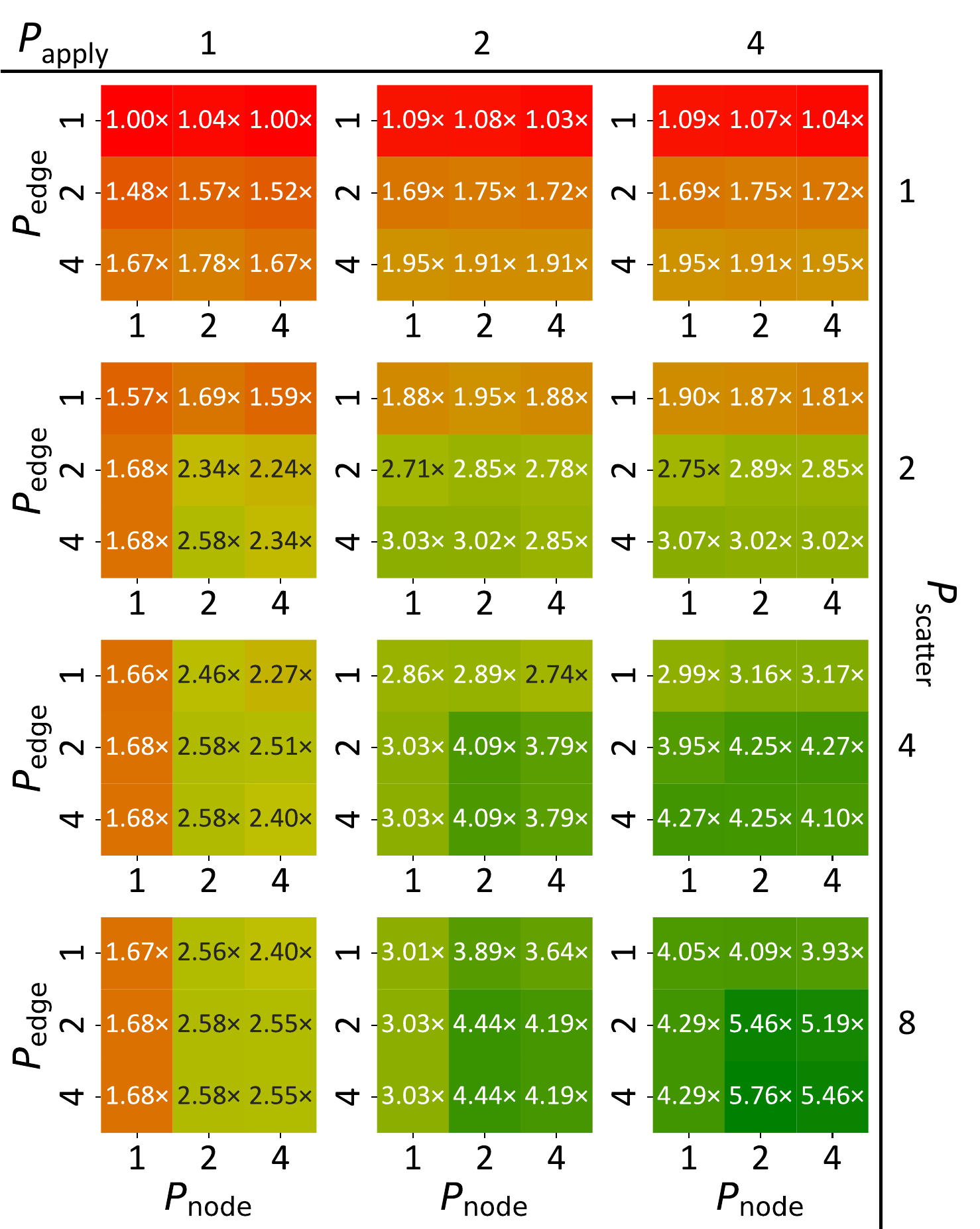}
    \caption{Overall speedup resulting from different parallelization factors $P_\text{node}$, $P_\text{edge}$, $P_\text{apply}$, and $P_\text{scatter}$.}
    \label{fig:parallel}
    \vspace{-12pt}
\end{figure}

We then evaluate the impact of varying {\gnnacc}'s configurable parallelism factors as a design space exploration (DSE). On top of the baseline architecture, we observe how varying $P_{\text{node}}$ from 1 to 4, $P_{\text{edge}}$ from 1 to 4, $P_{\text{apply}}$ from 1 to 4, and $P_{\text{scatter}}$ from 1 to 8 change the overall inference latency.
We use GCN model on MolHIV dataset.
We plot 108 data points in Fig.~\ref{fig:parallel}, showing the speedup over the baseline, where all four parameters are 1.
\underline{First}, each 3$\times$3 block evaluates the impact of edge/node parallelism. Increasing $P_{\text{node}}$ and $P_{\text{edge}}$ usually lead to speedup, from 1.33$\times$ to 1.84$\times$ when both parameters increase from 1 to 4.
In some cases, increasing one parameter does not affect latency, depending on the bottleneck. %
When $P_{\text{apply}}$ = 1, $P_{\text{scatter}}$ = 2, and $P_{\text{node}}$ = 1, increasing $P_{\text{edge}}$ from 2 to 4 has no effect, but increasing $P_{\text{node}}$ to 2 improves speedup from 1.68$\times$ to 2.34$\times$. Here, NT is the bottleneck, not MP.
\underline{Second}, increasing $P_{\text{apply}}$ and $P_{\text{scatter}}$ always seems effective%
, from 1.09$\times$ to 2.99$\times$ when both increase from 1 to 4.
The largest speedup within the design space is 5.76$\times$, when $P_{\text{edge}}$ = 4, $P_{\text{node}}$ = 2, $P_{\text{apply}}$ = 4, and $P_{\text{scatter}}$ = 8.

On one hand, the DSE study demonstrates that increasing the four parallelism parameters can effectively reduce latency. On the other hand, we recognize that the speedup is not linear, e.g., doubling one parameter does not result in 2$\times$ speedup. One reason is the entangled effect of the four parameters, where some are bottlenecks while others are not; another reason is imbalanced MP workload due to imbalanced graphs.

\subsection{Workload Imbalance Analysis}

As edges are assigned to MP units by target node ID, workload imbalance may result. In Table~\ref{tab:workload-imbalance}, we analyze workload imbalance for varying values of \(P_\text{edge}\), given by the largest difference in workloads between any two MP units as a percentage of the total workload. We see no more than 8.82\% imbalance for all evaluated combinations of dataset and \(P_\text{edge}\); nevertheless, we will consider improvements in future work.

\begin{table}
    \centering
    \caption{Workload imbalance for varying values of \(P_\text{edge}\). At 0\%, all MP units have identical workloads; at 100\%, one unit handles the entire workload.}
    \footnotesize
    \renewcommand{\arraystretch}{0.9}
    \vspace{-8pt}
    \setlength{\tabcolsep}{3pt}
    \begin{tabular}{c|c|c|c|c|c|c|c}
    \toprule
    & \multicolumn{7}{c}{\textbf{Datasets}} \\
    \(\boldsymbol{P}_\text{\textbf{edge}}\) & MolHIV & MolPCBA & HEP & Cora & CiteSeer & PubMed & Reddit \\
    \midrule
    2 & 6.41\% & 5.58\% & 2.47\% & 0.95\% & 0.40\% & 0.41\% & 0.04\% \\
    4 & 8.59\% & 7.78\% & 3.24\% & 3.83\% & 1.67\% & 2.21\% & 0.17\% \\
    8 & 8.82\% & 7.82\% & 3.30\% & 2.56\% & 2.69\% & 1.81\% & 0.28\% \\
    16 & 8.34\% & 7.62\% & 3.12\% & 2.72\% & 2.36\% & 1.23\% & 0.21\% \\
    32 & 7.37\% & 6.25\% & 3.75\% & 1.95\% & 1.68\% & 0.87\% & 0.21\% \\
    64 & 7.27\% & 6.28\% & 3.95\% & 1.82\% & 1.22\% & 0.82\% & 0.16\% \\
    \bottomrule
    \end{tabular}
    \label{tab:workload-imbalance}
\end{table}

\subsection{Comparison Against GCN Accelerators}
\label{sec:gcn_compare}

\begin{table}
\vspace{-4pt}
    \centering
    \caption{Comparison with I-GCN~\cite{igcn} and AWB-GCN~\cite{awbgcn} on the Cora, CiteSeer, PubMed, and Reddit datasets using GCN. Although not a fair comparison, we still achieve an average of 1.26$\times$ normalized speedup and an average of 1.55$\times$ energy efficiency (EE) over I-GCN.}
    \vspace{-8pt}
    \footnotesize
    \setlength{\tabcolsep}{3pt}
    \renewcommand{\arraystretch}{0.9}
    \begin{tabular}{c|c|c|c|c|c}
        \toprule
        Dataset & Accelerator & \begin{tabular}{@{}c@{}} Latency \\ (\textmu s) \end{tabular} & DSPs & \begin{tabular}{@{}c@{}} Normalized \\ by DSPs \end{tabular} & \begin{tabular}{@{}c@{}} EE \\ (graph/kJ) \end{tabular}\\
        \midrule
        \multirow{3}{*}{Cora} & AWB-GCN & 2.3 & 4096 & 2.3 & 3.1{\scriptsize E}6 \\
        & I-GCN & 1.3 & 4096 & 1.3 & 7.1{\scriptsize E}6 \\
        & \textbf{{\gnnacc}} & 6.912 & 747 & \textbf{1.261 (1.03\texttimes{})} & \textbf{7.77{\scriptsize E}6 (1.09\texttimes{})} \\
        \midrule
        \multirow{3}{*}{CiteSeer} & AWB-GCN & 4.0 & 4096 & 4.0 & 1.9{\scriptsize E}6 \\
        & I-GCN & 1.9 & 4096 & 1.9 & 3.7{\scriptsize E}6 \\
        & \textbf{{\gnnacc}} & 8.332 & 747 & \textbf{1.520 (1.25\texttimes{})} & \textbf{6.44{\scriptsize E}6 (1.74\texttimes{})} \\
        \midrule
        \multirow{3}{*}{PubMed} & AWB-GCN & 30 & 4096 & 30 & 2.5{\scriptsize E}5 \\
        & I-GCN & 15.1 & 4096 & 15.1 & 5.3{\scriptsize E}5 \\
        & \textbf{{\gnnacc}} & 53.22 & 747 & \textbf{9.706 (1.56\texttimes{})} & \textbf{1.01{\scriptsize E}6 (1.91\texttimes{})} \\
        \midrule
        \multirow{3}{*}{Reddit} & AWB-GCN & 3.2{\scriptsize E}4 & 4096 & 3.2{\scriptsize E}4 & 2.1{\scriptsize E}2 \\
        & I-GCN & 3.0{\scriptsize E}4 & 4096 & 3.0{\scriptsize E}4 & 3.5{\scriptsize E}2 \\
        & \textbf{{\gnnacc}} & 1.36{\scriptsize E}5 & 747 & \textbf{2.49{\scriptsize E}4 (1.20\texttimes{})} & \textbf{3.94{\scriptsize E}2 (1.13\texttimes{})} \\
        \bottomrule
    \end{tabular}
    \label{tab:gcn_compare}
    \vspace{-12pt}
\end{table}

As discussed earlier, most existing GNN accelerators formulate GNNs as GEMM/SpMM and cannot handle advanced features such as edge embeddings. I-GCN \cite{igcn} and AWB-GCN \cite{awbgcn} are two works of this type. In this section, we implement the GCN architecture of these existing works within the {\gnnacc} framework and compare its performance on the Cora, CiteSeer, PubMed, and Reddit datasets.
We follow their model configuration, using a two-layer GCN, with node embedding dimension being 16 and no edge embedding.

Results are shown in Table~\ref{tab:gcn_compare}. Given the different hardware platform, we normalize by the number of DSPs, the major computation resource, and %
achieve average performance 1.26$\times$ faster than I-GCN with 1.55$\times$ energy efficiency, which is 2.21$\times$ faster and 2.95$\times$ more energy efficient than AWB-GCN.
Although not a fair comparison because the optimizations in I-GCN and AWB-GCN do not generalize to edge embeddings, we still show superior performance.

\section{Conclusion}

We proposed \textbf{\gnnacc}, the first generic and flexible accelerator framework for a wide range of GNNs.
It uses a novel dataflow architecture with multiple levels of parallelism, including edge/node and apply/scatter parallelism.
The enabling techniques are multiple data queues with on-the-fly node multicasting.
Its noteworthy features include generality for future-proofing, real-time processing, and open-source.
On-board evaluation with guaranteed functionality exhibited invariant speed-up comparing with CPU and GPU, up to 24--254$\times$ against CPU and 1.3--477$\times$ against GPU with batch sizes from 1024 to 1.
We are also 1.26$\times$ faster and 1.55$\times$ more energy efficient than the SOTA GNN accelerator.
Remarkable speedup suggests that we did not sacrifice performance for generality and can deliver real-time performance.

\section*{Acknowledgements}
The authors would like to thank Zihang Qiao for his contributions to the development of PNA and DGN, as well as Parima Mehta for her contribution to the virtual node model.
The authors would also like to thank Dr. Pan Li for his insightful discussions.

\bibliographystyle{unsrt}
\bibliography{refs}

\end{document}